\documentclass[useAMS,usenatbib,referee]{mnras}

\usepackage{lineno}
\modulolinenumbers[1]
\usepackage{hyperref}
\usepackage{amsmath}
\usepackage{amssymb}

\usepackage{psfrag} 

\usepackage[utf8]{inputenc}
\usepackage{graphicx}
\usepackage{txfonts}
\usepackage[T1]{fontenc}

\usepackage{relsize}
\usepackage{pbox}
\usepackage{bm}

\usepackage[svgnames]{xcolor}
\usepackage{soul}

\newcommand{\Eqref}[1]{Eq.\,(\ref{#1})}
\newcommand{\eqsref}[1]{Eqs.\,(\ref{#1})}
\newcommand{\figref}[1]{Fig.\,\ref{#1}}
\newcommand{\tabref}[1]{Tab.\,\ref{#1}}
\newcommand{\secref}[1]{Sec.\,\ref{#1}}

\newcommand{\inst}[1]{$^{#1}$}

\definecolor{DarkBrown}{rgb}{.396,.263,.129}
\definecolor{LightBrown}{rgb}{.698,.463,.227}
\definecolor{darkgreen}{rgb}{0.0,.4,0.0}

\newcommand{\CFL}{C_{\mathrm{CFL}}}

\title[Anomalous dynamics triggered by a non-convex EoS in relativistic flows]%
{
Anomalous dynamics triggered by a non-convex equation of state in relativistic flows
}

\author[NCT~et~al.]
{
J.M.\,Ib{\'a}\~nez\inst{1},
  A.\,Marquina\inst{2},
  S.\,Serna\inst{3},
  M.A.\,Aloy\inst{1}  
\\
\inst{1} Departamento de Astronom\'{\i}a y Astrof\'{\i}sica,  Universidad de Valencia,  C/ Dr.~Moliner 50, 46100 Burjassot, Spain  \\
\inst{2} Departamento de Matem\'aticas,  Universidad de Valencia,  C/ Dr.~Moliner 50, 46100 Burjassot, Spain\\
\inst{3} Departament de Matematiques,  Universitat Autonoma de Barcelona, 08193 Bellaterra-Barcelona, Spain\\
}
\begin{document}

\date{Accepted 2018 January 5. Received 2017 December 8; in original
  form 2017 September 14}

\pagerange{\pageref{firstpage}--\pageref{lastpage}} \pubyear{2018}

\maketitle

\label{firstpage}

\begin{abstract}
  The non-monotonicity of the local speed of sound in dense matter at
  baryon number densities much higher than the nuclear saturation
  density ($n_0 \approx 0.16\,$fm$^{-3}$) suggests the possible
  existence of a non-convex thermodynamics which will lead to a
  non-convex dynamics. Here, we explore the rich and complex dynamics
  that an equation of state (EoS) with non-convex regions in the
  pressure-density plane may develop as a result of genuinely
  relativistic effects, without a classical counterpart. To this end,
  we have introduced a phenomenological EoS, whose parameters can be
  restricted heeding to causality and thermodynamic stability
  constraints. This EoS shall be regarded as a toy-model with which we
  may mimic realistic (and far more complex) EoS of practical use in
  the realm of Relativistic Hydrodynamics.
\end{abstract}
\begin{keywords}
  relativistic processes -- shock waves -- dense matter -- equation of state --
    hydrodynamics -- methods:numerical
\end{keywords}


\section{Introduction}
\label{sec:intro}

The complexity of the numerical models based upon a fluid description
of the physical constituents (i.e., based upon a hydrodynamic or
magnetohydrodynamic modeling) has grown pace to pace with the
sustained increase of computational power available to the scientific
and engineering community.  Among other factors, the equation of state
(EoS) employed sets the degree of realism with which physical systems
are modeled. The EoS is a constitutive relation, which provides the
closure to the sets of balance laws that account for the conservation
of various basic quantities of the system (e.g., mass, momentum,
energy, magnetic flux, etc.).  There has been a progressive shift from
assuming that the underlying fluid was governed by a simple EoS with,
e.g., a constant adiabatic index, $\gamma$, to a more realistic
microphysical description, where a general EoS, allowing for thermal
and/or chemical processes, is decisive to shape both the
thermodynamics and the dynamics of the flows. Numerous are the fields
in which a complex EoS is needed. For instance, in Astrophysics, the
treatment of the interstellar medium in the process of fragmentation
of self-gravitating molecular clouds
\citep[e.g.,][]{Spaans:2000ApJ...538..115S,Li:2003ApJ...592..975L,Jappsen:2005A&A...435..611J};
the thermo-chemical evolution of the primordial stars
\citep[e.g.,][]{Yoshida:2006ApJ...652....6Y,Glover:2008MNRAS.388.1627G,Clark:2011ApJ...727..110C,Greif:2011ApJ...737...75G};
the envelopes of young planets embedded in protoplanetary disks
\citep{DAngelo:2013ApJ...778...77D,Roth:2015ApJS..217....9R}; the
spectral modeling of stellar atmospheres
\citep[e.g.,][]{Asplund:1999A&A...346L..17A,Collet:2007A&A...469..687C};
the stellar evolution
\citep[e.g.,][]{Kippenhahn:1990stellar,Rauscher:2002ApJ...576..323R};
the ionization of cold interstellar neutral gas as shock waves heat it
\citep[e.g.,][]{Flower:2003,Vaidya:2015A&A...580A.110V}, etc. However,
complex EoSs are not restricted to astrophysics or plasma
physics. They are also common in material processing, industrial
frameworks and in the study of dense gas near the liquid-vapor
saturation curve \citep{Menikoff:2007, Thompson:1973JFM}.

The convexity of any EoS is mathematically defined in terms of the
value of the fundamental derivative, $\mathcal G$ (see
Sect.~\ref{sec:PhenoEoS}). The fundamental derivative measures the
convexity of the isentropes in the $p-\rho$ plane (where $p$ is the
pressure and $\rho$ the rest-mass density). If ${\cal G} > 0$,
isentropes in the $p-\rho$ plane are convex, leading to {\em
  expansive} rarefaction waves and {\em compressive} shocks
\citep{Thompson:1971PhFl...14.1843T, Rezzolla:2013rehy.book}. This is
the usual regime in which many astrophysical scenarios
develop. However, some EoSs may display regimes in which
${\cal G} < 0$, i.e., the EoS is non-convex. The non-convexity of
isentropes in the$p-\rho$ plane yield, e.g., {\em compressive}
rarefaction waves and {\em expansive} shocks. These non-classical or
{\em exotic} phenomena have been observed experimentally
\citep{Cinella:2007JFM...580..179C,Cinella:2011PhFl...23k6101C}.

Prototype cases described by an extremely complex EoS involve systems
where matter is so compact that its density is close to the nuclear
saturation density ($n_0 \approx 0.15-0.16$\,fm$^{-3}$).  The
properties of the EoS for dense matter have a crucial influence in
many different problems in Astrophysics and Nuclear Physics, some of
which are: (1) the equilibrium and dynamics of compact stars
\citep[e.g.,][]{Baumgarte:2000ApJ...528L..29B,Hempel:2012ApJ...748...70H,Banik:2014ApJS..214...22B},
(2) the merger of compact objects
\citep[e.g.,][]{Kiuchi:2009PhRvD..80f4037K,Hotokezaka:2011PhRvD..83l4008H,Bauswein:2014PhRvD..90b3002B,Takami:2014PhRvL.113i1104T},
(3) the evolution of the early Universe
\citep[e.g.,][]{Munyoz:2015,Borsanyi:2016}, (4) the collision of heavy
ions \citep[e.g.,][]{Laine:2006,Nonaka:2007,Luzum:2008}, etc.  In
spite of the important efforts (theoretical and experimental) the
nuclear physicists and astrophysicists communities are still far from
having an accurate knowledge of the properties of the EoS for dense
matter (i.e., a fundamental physical issue).

At densities much higher than $n_0$ nuclear/hadronic matter undergoes
a transition into a quark-gluon plasma (QGP). The nature of the
finite-temperature QCD transition (first-order, second-order or
analytic crossover) remains ambiguous
\citep{Aoki:2006Natur.443..675A}.  Two groups, the HotQCD
Collaboration \citep{Bazazov:2014PhRvD..90i4503B} and the
Wuppertal-Budapest Collaboration \citep{Borsanyi:2014PhLB..730...99B},
have reported results - using QCD lattice techniques - about the EoS
of the QGP that characterizes the transition from the hadronic phase
into the QGP phase. Their findings in the continuum extrapolated EoS
favor the crossover nature of the transition. In the
phenomenologically relevant range of temperature, $130 - 400\,$MeV,
the results from both groups show similarities with regard to the
trace anomaly, pressure, energy density and entropy density.  The
energy density in the crossover region,
$145\le T \text{(MeV)} \le 163$, is a factor of about $(1.2 - 3.1)$
times the nuclear saturation density. Remarkably, within the former
temperature range the square of the sound speed, $c_{\rm s}^2$, is
non-monotonic, displaying a minimum at
$T_{c^2_{\rm s, min}} \approx 145 -150\,$MeV. From that minimum and up
to $T\simeq 200$\,MeV, $c_{\rm s}^2<c^2/3$, where $c$ is the speed of
light in vacuum. Following a different strategy, which combines the
knowledge of the EoS of hadronic matter at low densities with the
observational constraints on the masses of neutron stars,
\cite{Bedaque:2015PhRvL.114c1103B} have concluded that the sound speed
of dense matter is a non-monotonic function of density.  In dense
matter, the non-monotonicity of the speed of sound can also result
from the behaviour of the adiabatic index (see definition below) in,
e.g., the Skyrme Lyon (SLy;
\citealt{Chabanat:1997NuPhA.627..710C,Chabanat:1998NuPhA.635..231C})
and in the Friedman Pandharipande Skyrme (FPS;
\citealt{Pandharipande:1989ASIB..205..103P}) EoSs (see Fig. 5 in
~\citealt{Haensel:2004A&A...428..191H} and \figref{fig:HP04}
below). More recently, \cite{Shen:2011PhRvC..83f5808S} have shown the
non-monotonic behaviour of the adiabatic index for various EoSs
broadly used in numerical simulations of core collapse, supernovae and
compact object mergers (see their Fig.\,9). Again, we remark that,
beyond the disputable dip in the adiabatic index below $n_0$, which is
typically attributed to the phase transition from non-uniform to
uniform matter, all the equations of state compared in
\cite{Shen:2011PhRvC..83f5808S} show that there is a range of
densities above $n_0$ such that the adiabatic index decreases with
density. Thus, the non-monotonicity of the adiabatic index (likewise,
of the sound speed) should be considered as a genuine feature of
matter at a few times the nuclear saturation density. Under such
conditions, and particularly if there are phase transitions to exotic
components, we advance that the fundamental derivative (see
Sect.~\ref{sec:PhenoEoS}) could be negative, implying that the EoS be
non-convex in that regime.


In Sec.\,\ref{sec:equations} we show the equations of special
relativistic hydrodynamics (SRHD) and recap their spectral properties
for the readers benefit. Though these concepts are not new, they serve
for the purpose of linking the convexity properties of the system of
SRHD equations to its characteristic structure and to the convexity of
the EoS. Studying the dynamics and thermodynamics of a realistic,
state-of-the-art, non-convex EoS in a transrelativistic regime is a
difficult task, in which obtaining any analytic insight is
doubtful. Thus, we propose to illustrate the potential effects that a
non-convex EoS may produce on both the thermodynamics and the dynamics
of relativistic models with a phenomenological (analytic and very
simple) EoS. This EoS has the virtue of mimicking the loss of
convexity resulting from a non-monotonic behaviour of the adiabatic
index with density, while at the same time be simple enough to gain a
deeper physical insight in a number of phenomena of interest
(Sect.~\ref{sec:PhenoEoS}). As we shall see, it is possible to tune
the parameters of this phenomenological EoS to emulate the behaviour
of the sound speed in very dense plasma (above nuclear matter
density).  The main focus of this work will be to carefully examine
the regime in which purely relativistic effects determine the
non-convex evolution of prototype relativistic Riemann problems of
flows obeying our phenomenological EoS (Sect.~\ref{sec:RPs}).  Since
the relativistic effects we discuss here are independent of the
numerical methods employed to solve the equations of SRHD, we defer
for a follow up paper the comprehensive study of the numerical
approximation of the non-convex special relativistic flows, governed
by the current EoS, through the design of novel shock capturing
schemes and a set of numerical experiments \citep{Marquina_etal:2017}.
Our main conclusions will be summarized in Sect.~\ref{sec:summary}. 

\section{The equations of SRHD}
\label{sec:equations}

The equations of SRHD
  correspond to the conservation of rest-mass and energy-momentum of a
  fluid. In a flat space-time and Cartesian coordinates, these
  equations read \cite[e.g.,][]{Font_etal:1994,Aloy_etal:1999}:
\begin{align}
\label{eq:cont}
J^\mu_{\,\,\,\,,\mu} = 0, \\
\label{eq:e-mom}
T^{\mu \nu}_{\,\,\,\,\,\,,\mu} = 0,
\end{align}
\noindent
where subscripts $(\,_{,\mu}\,)$ denote the partial derivative with
respect to the corresponding coordinate,
$(t,x,y,z)$\footnote{Throughout this paper, Greek indices will run
  from 0 to 3, while Roman run from 1 to 3, or, respectively, from $t$
  to $z$ and from $x$ to $z$, in Cartesian coordinates.}, and the
standard Einstein sum convention is assumed.  $J^\mu$ and
$T^{\mu \nu}$ are the four-current density of rest mass and the
energy-momentum tensor, respectively,
\begin{align}
J^\mu &= \rho u^\mu, \\
T^{\mu \nu} &= \rho h u^\mu u^\nu + \eta^{\mu \nu} p.
\end{align}
$\rho$ is the proper rest-mass density, $h =1 + \varepsilon + p/\rho$
is the specific enthalpy, $\varepsilon$ is the specific internal
energy, $p$ is the thermal pressure, and $\eta^{\mu \nu}$ is the
Minkowski metric of the space-time where the fluid evolves. Throughout
the paper we use units in which the speed of light is $c=1$. The
four-vector representing the fluid velocity, $u^\mu$, satisfies the
condition $u^\mu u_\mu = -1$.

The system formed by \eqsref{eq:cont} and \eqref{eq:e-mom} can be
written in explicit conservation form as follows
\begin{equation}
\frac{\partial {\bf U}}{\partial t} +
\frac{\partial {\bf F}^{i}}{\partial x^{i}} = 0
\label{eq:system}
\end{equation}
\noindent
where ${\bf V} = (\rho, v^j, \epsilon)^T$ is the set of primitive
variables.  The state vector (the set of conserved variables) ${\bf U}$
and the fluxes, ${\bf F}^i$, are, respectively:
\begin{eqnarray}
{\bf U} & = & \left(\begin{array}{c}
  D    \\
  S^j  \\
  \tau \\
\end{array}\right),
\label{state_vector}
\end{eqnarray}
\begin{eqnarray}
{\mathbf F}^i & =& \left(\begin{array}{c}
  D v^i \\
  S^j v^i + p \delta^{ij} \\
  (\tau + p) v^i
\end{array}\right).
\label{flux2}
\end{eqnarray}

In the preceding equations, $D$, $S^j$ and $\tau$ stand, respectively,
for the rest-mass density, the momentum density of the magnetized
fluid in the $j$-direction, and its total energy density, all of them
measured in the laboratory (i.e., Eulerian) frame:
\begin{align}
\label{eq:D}
  D &= \rho W,\\
\label{eq:Sj}
  S^j &= \rho h W^2 v^j,\\
\label{eq:tau}
  \tau &= \rho h W^2 - p - D.
\end{align}
The components of the fluid three-velocity, $v^i$, as measured in the
laboratory frame, are related with the components of the fluid
four-velocity through $u^\mu = W(1, v^i)$, where $W$ is the flow
Lorentz factor, $W=1/\sqrt{(1-v^i v_i)}$.

Recapping the arguments of \cite{Ibanez:2013CQGra..30e7002I} , the set
of five equations \Eqref{eq:system} form a nonlinear, hyperbolic
system of conservations laws (HSCL), since given an arbitrary linear
combination of the Jacobians of the system,
$\partial {\bf F}^{i} / \partial {\bf U}$, there exist five
eigenvectors $\{{\bf r}_p\}_{p=1,\ldots,5}$, each associated to its
corresponding eigenvalue, $\lambda_p$, which form a basis of
$\mathbb{R}^5$.  Among the characteristic fields ${\cal C}_p$ of
system (\ref{eq:system}) satisfying
\begin{equation}
{\cal C}_p:\: \frac{d{\bf x}}{dt}=\lambda_p,
\end{equation}
some can be linearly degenerate, i.e.,
\begin{equation}
{\cal P}_p := {\bf \nabla}_{\bf U}\lambda_p \cdot {\bf r}_p = 0,
\label{eq:linearly_denegerated}
\end{equation}
where ${\bf \nabla}_{\bf U}\lambda_p$ is the gradient of
$\lambda_p({\bf U})$ in the space of conserved variables and the dot
stands for the inner product in $\mathbb{R}^5$.  Contact
discontinuities are the only admissible discontinuities along a
linearly degenerated field. Likewise, there may exist characteristic
fields, which are genuinely nonlinear, i.e.,
\begin{equation}
{\cal P}_p := {\bf \nabla}_{\bf U}\lambda_p \cdot {\bf r}_p \ne 0.
\label{eq:genuinely_nonlinear}
\end{equation}
Along a genuinely nonlinear characteristic field shocks (but not
contact discontinuities) may develop.

A HSCL is said to be convex if all its characteristic fields are
either genuinely nonlinear or linearly degenerate. In a non-convex
system, non-convexity is associated with those states ${\bf U}$ for
which one ${\cal P}_p$ is zero and changes sign in a neighborhood of
${\bf U}$. The convexity of the HSCL ultimately depends on the EoS,
since the value of the dot products defined in
\eqsref{eq:linearly_denegerated} and \eqref{eq:genuinely_nonlinear}
depend on the relations among the thermodynamic variables. For an
ideal gas EoS with constant adiabatic index, the HSCL formed by the
SRHD equations is convex. This is not necessarily guaranteed for other
EoSs of interest in Astrophysics and Nuclear Physics. 
\section{A phenomenological EOS}
\label{sec:PhenoEoS}

The system of SRHD equations (Sec.\,\ref{sec:equations}) must be
closed with a suitable EoS, i.e., a relation between thermodynamic
variables of, e.g., the form $p=p(\rho,\varepsilon)$. In order to
explore departures from the regular convex behaviour ubiquitously
found in Relativistic Astrophysics, we propose that the pressure obeys
an ideal gas-like EoS of the form:
\begin{equation}
p = (\gamma - 1) \rho \varepsilon + B \rho , 
\label{eq:p(rho)}
\end{equation}
where
\begin{equation}
\gamma  =  \gamma_0 + {\cal K} \exp{(-x^2/\sigma^2)}
\label{eq:Gaussianterm}
\end{equation}
and
\begin{equation}
{\cal K} =  \gamma_1 - \gamma_0 \,,\, x  =  \rho - \rho_1 
\end{equation}
being $\gamma_1 = \gamma(\rho_1)$, and $\rho_1$ plays (in the
exponential) the role of a simple scale factor for the density.
Henceforth, the phenomenological EoS (\Eqref{eq:p(rho)}) encompassing
a Gaussian gamma-law plus a linear term in density will be named
'GGL'. The free parameters of the GGL EoS are:
$\gamma_0, \gamma_1, \rho_1, \sigma$ and $B$ (see
Sec.\,\ref{sec:typicalvalues} for typical values)\footnote{$\sigma$,
  in our GGL EoS, is related to the Gaussian FWHM, since
  $2\sqrt{\ln{2}} \sigma \approx 1.665 \sigma$}. Since the GGL EoS
depends on five parameters, we may employ this freedom in order to
mimic existing features in other realistic EoS. Furthermore, we may
adjust them to obtain a non-convex thermodynamics.  To explore this
possibilities we derive a number of thermodynamic quantities
corresponding to the GGL EoS.

The GGL EoS contains two obvious differences with respect to an
standard ideal gas EoS with constant adiabatic index. Namely, the
Gaussian, rest-mass density dependent term (\Eqref{eq:Gaussianterm})
and the linear term $B\rho$. The Gaussian term is introduced to
emulate the non-monotonic dependence of the adiabatic index with the
rest-mass density exhibited by most nuclear matter EoSs (see,
\secref{sec:local_sound}). The motivation for including the linear
term $B\rho$ in \Eqref{eq:p(rho)} is simple. In the high-density
limit, $\rho \gg \rho_1$, the Gaussian term in \Eqref{eq:Gaussianterm}
vanishes. A value $B=\gamma_0-1$ results in a barotropic EoS
\citep{Anile:2005,Ibanez:2012CQGra..29o7001M} in the high-density
regime, since the pressure depends only upon the total mass-energy
density $e=\rho(1+\varepsilon)$; $p=p(e)$. Specifically, we find
$p=Be$.

\subsection{Local sound velocity and adiabatic index}
\label{sec:local_sound}

The classical speed of sound,
$\displaystyle{
c_{\rm s_{(C)}}
:= \sqrt{\left.\frac{\partial p}{\partial \rho} \right|_{s}}}$,
for our GGL EoS, is
\begin{equation}  
c_{\rm s_{(C)}}^2  = 
\displaystyle{
\gamma \, \left(\frac{p}{\rho} \,\,+\,\,\varepsilon
\,\,\frac{d\ln{\gamma}}{d\ln{\rho}} \right) 
}  
 = 
\displaystyle{
\gamma \,\, \left\{\varepsilon \left(\gamma - 1 + \frac{d\ln{\gamma}}{d\ln{\rho}} \right) \, + \, B \right\}
}.
\end{equation}  
%
%

The relativistic definition of the speed of sound is related with the classical one by
the specific enthalpy, $h = 1 + \varepsilon + p/\rho$, according to the following expression
\begin{equation}                                                                  
c_{\rm s_{(C)}}^2  =  h  c_{\rm s_{(R)}}^2.
\end{equation}                                                                  
For our GGL EoS the specific enthalpy reads 
\begin{equation}                                                                  
 h   =   1   +   \gamma \varepsilon   +   B .
\end{equation}                                                                  


The adiabatic index \citep[$\Gamma_1$, see,
e.g.,][]{Chandrasekhar:1939isss.book} of the GGL EoS is
\begin{equation}
\Gamma_1 := \displaystyle{\left.\frac{\partial \ln{p}}{\partial \ln{\rho}} \right|_{s} } =
\displaystyle{\left(\frac{\rho}{p}\right) c_{\rm s_{(C)}}^2} =
\displaystyle{
\gamma \, 
\left(1 \,\,+\,\,\left(\frac{\rho \varepsilon}{p}\right) \,\,\frac{d\ln{\gamma}}{d\ln{\rho}} \right) 
}.
\end{equation}

\begin{figure*}
\centering
\includegraphics[width=0.45\textwidth]{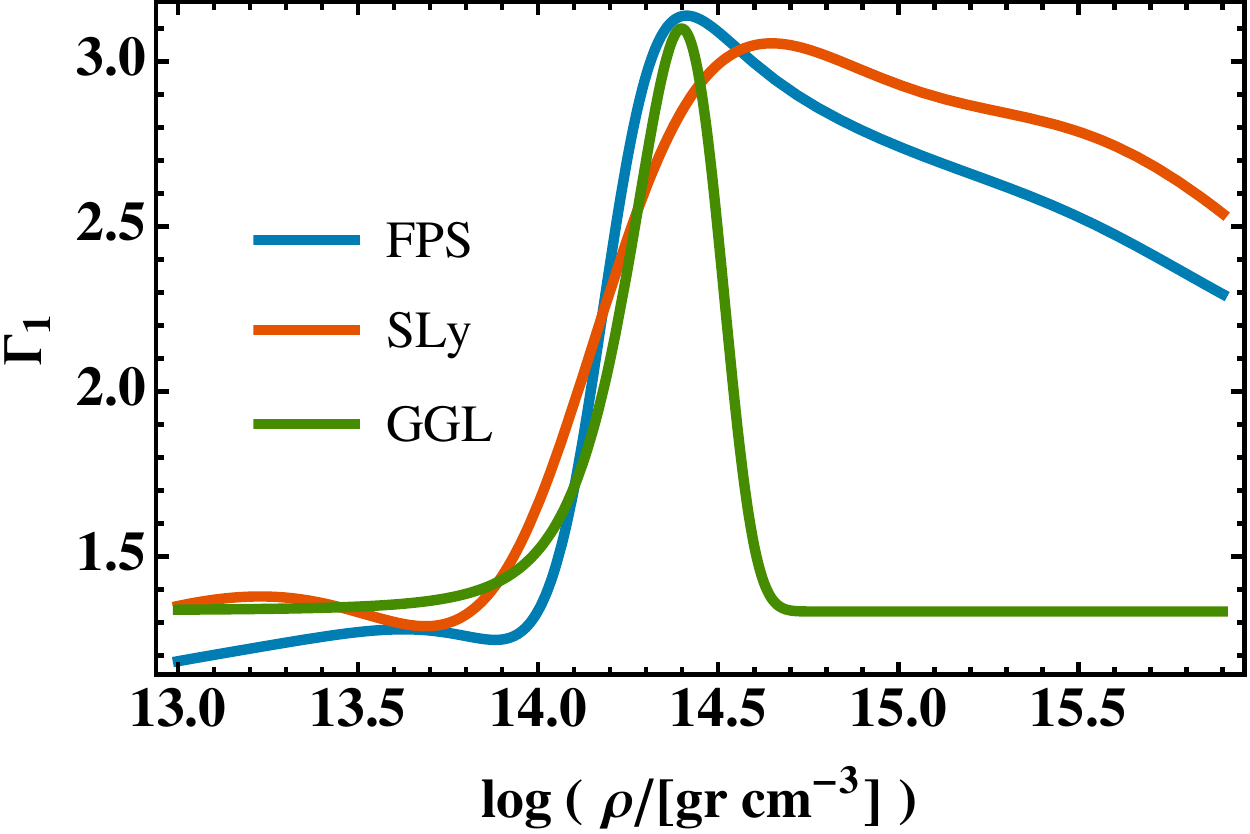} 
\includegraphics[width=0.45\textwidth]{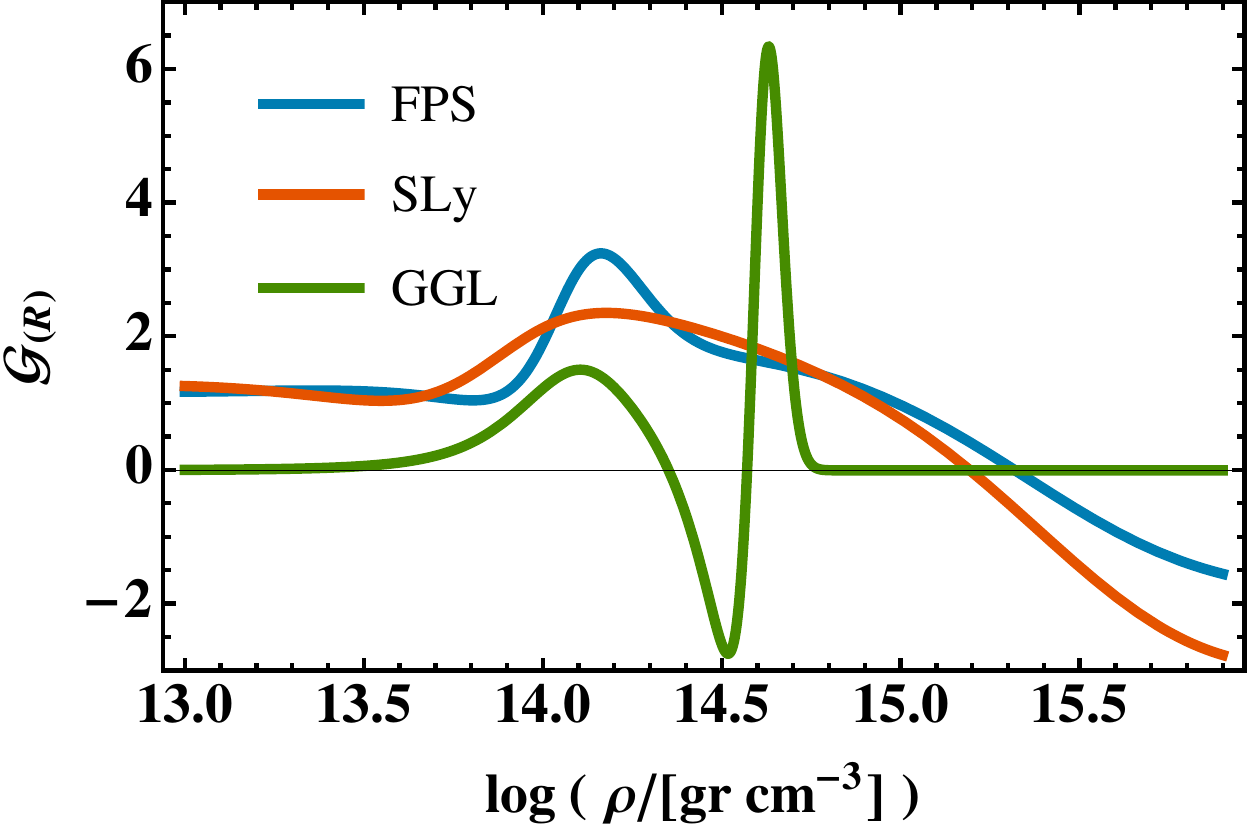}

\caption{Adiabatic index (left panel) and relativistic fundamental
  derivative (right panel) as a function of the logarithm of the
  rest-mass density for the SLy, FPS and GGL EoSs. To obtain a
  qualitatively similar behaviour of the GGL EoS as in the case of the
  other two nuclear matter EoSs, we set $\gamma_0 = 4/3$,
  $\gamma_1 = 3.1$, $\rho_1 = 2.5\times 10^{14}\,$gr\,cm$^{-3}$,
  $\sigma = 0.4$, $\varepsilon=0.01c^2$ and $B=\gamma_0-1$. We note
  that, for $B=0$ the adiabatic index is independent of $\varepsilon$
  and of $p$.}
\label{fig:HP04}
\end{figure*}

We note that, in general, $\Gamma_1 \,\,\ne \,\, \gamma$. $\Gamma_1$
is an important dimensionless parameter characterizing the stiffness
of the EoS at given density. As we anticipated in the introduction,
the non-monotonic behaviour of the adiabatic index is a hint that
points towards a non-convex thermodynamics in nuclear matter
EoSs. Actually, our phenomenologic GGL EoS finds its motivation in
reproducing the local maximum in the adiabatic index resulting from
the stiffening of the EoS above nuclear matter density.  In
\figref{fig:HP04} (left panel) we show the variation of the adiabatic
index with the rest-mass density in a range relevant for stellar core
collapse. The adiabatic index of the SLy
\citep{Chabanat:1997NuPhA.627..710C,Chabanat:1998NuPhA.635..231C} and
FPS \citep{Pandharipande:1989ASIB..205..103P} EoSs has been computed
using the analytic fits of \cite{Haensel:2004A&A...428..191H}. Tuning
the parameters of our phenomenological approximation (see caption of
\figref{fig:HP04}), we observe that the GGL EoS captures properly the
abrupt increase of $\Gamma_1$ for
$\rho\gtrsim 10^{14}\,$gr\,cm$^{-3}$. The decay post maximum is much
faster in the GGL EoS than in the two realistic EoS considered
here. However, the decay rate of $\Gamma_1$ after the maximum changes
from one to another realistic EoSs. For instance, the SLy EoS displays
a shallower fall-off of the adiabatic index beyond the maximum than
the FPS EoS. We have not incorporated an additional parameter to allow
for the asymmetric behavior of $\Gamma_1$ around the maximum in the
GGL EoS for simplicity. For comparison, we point out that the analytic
fits of \cite{Haensel:2004A&A...428..191H} require of, at least, 18
different tunable parameters. Thus, it is not surprising that the
thermodynamic quantities computed with the GGL EoS display large
deviations with respect to any existing EoS. This may specially happen
far from the regions in which an specific set of the (five) parameters
of our GGL EoS have been tuned to follow more closely the behavior of
a given nuclear matter EoSs.  We insist on the fact that the purpose
of the GGL EoS is not to substitute any realistic EoS but, instead,
exemplify the potential occurrence of non-convex regions in EoS
commonly used in Astrophysics applications.

\subsection{The classical and the relativistic fundamental derivatives}
\label{sec:fundamentalderivatives}

  In classical fluid dynamics, the convexity of the system is
determined by the EoS ~\citep{Menikoff:1989RvMP...61...75M,Godlewski:1996}  and, more specifically, by the 
so-called {\it fundamental derivative}, ${\mathcal G}_{\rm (C)}$ 
(see its definition - classical -  and properties in, e.g.,~\citealt{Thompson:1971PhFl...14.1843T,Menikoff:1989RvMP...61...75M,Guardone:2010JFM...642..127G})
\begin{equation}
{\mathcal G}_{\rm (C)}
:= - \frac{1}{2} \,V
\,\displaystyle{\frac{\displaystyle{\left.\frac{\partial^2 p}{\partial
          V^2}\right|_s}}{\displaystyle{\left.\frac{\partial
          p}{\partial V}\right|_s}}},
\label{G1}
\end{equation}
where $V:= 1/\rho$ is the specific volume and $s$ the specific entropy. Alternative expressions for ${\mathcal G}_{\rm
  (C)}$ are~\cite{Menikoff:1989RvMP...61...75M}
\begin{equation}
{\mathcal G}_{\rm (C)}
= \displaystyle{
\frac{1}{2} \left( 1 + \Gamma_1 + \left.\frac{\partial  \ln \Gamma_1}{\partial \ln \rho} \right|_s \right),
}
\label{G2}
\end{equation}
\noindent
and
\begin{equation}
{\mathcal G}_{\rm (C)}
= \displaystyle{ 1 + \left.\frac{\partial \ln{c_{\rm s_{(C)}}}}{\partial \ln{\rho}} \right |_s},
\label{G3}
\end{equation}

The fundamental derivative measures the convexity of the isentropes in
the $p-\rho$ plane and if ${\mathcal G}_{\rm (C)} > 0$ then the
isentropes in the $p-\rho$ plane are convex, and the rarefaction waves
are expansive.

Analogously to \Eqref{G2}, let us define ${\mathcal G}^+$ 
\begin{equation}
{\mathcal G}^+ := \displaystyle{
\frac{1}{2} \left\{ 1 + \gamma + \left(\frac{d\ln{\gamma}}{d\ln{\rho}} \right) \right\}}.
\end{equation}

From \Eqref{G3} and for our GGL EoS we obtain:
\begin{equation}
{\mathcal G}_{\rm (C)} = 
{\mathcal G}^+  + \displaystyle{\frac{\gamma}{2 c_{\rm s_{(C)}}^2 }
\left( (\gamma \varepsilon + B) \frac{d \ln{\gamma}}{d\ln{\rho}} 
+ \varepsilon \frac{d^2 \ln{\gamma}}{d (\ln{\rho})^2} \right) 
}
\end{equation}


\cite{Ibanez:2013CQGra..30e7002I} found that there exist a quantity
analogous to the classical fundamental derivative, which they call
relativistic fundamental derivative. The following relationship
between the classical and relativistic fundamental derivatives holds:
\begin{equation}                                                                
{\mathcal G}_{\rm (R)} = {\mathcal G}_{\rm (C)} - \frac{3}{2} \,\,c_{\rm s_{(R)}}^2 
\label{eq:G6}
\end{equation} 
An important consequence of the above result is the following: for a
non-convex EoS, there will be a region (in the $p-\rho$ plane) where
the thermodynamics is non-convex from the relativistic point of view,
i.e., there where ${\mathcal G}_{\rm (R)} \le 0$, but it is convex
from the classical point of view ${\mathcal G}_{\rm (C)} \ge 0$.

We show in the right panel of \figref{fig:HP04} the relativistic
fundamental derivative for the GGL EoS as well as for the analytic
fits of the FPS and SLy EoSs given in
\cite{Haensel:2004A&A...428..191H}. We observe that for the chosen
parameters of the GGL EoS (see caption of \figref{fig:HP04}), the
relativistic fundamental derivative becomes negative in coincidence
with decay post-maximum in $\Gamma_1$ (\figref{fig:HP04} left panel),
signalling the existence of a non-convex thermodynamic region
extending between $14.3\lesssim \log{\rho} \lesssim 14.6$. Since the
fall-off of $\Gamma_1$ in the FPS and SLy EoSs is not so steep, these
EoSs do not develop a non-convex region in the same density range as
the GGL EoS, yet ${\mathcal G}_{\rm (R)}$ drops in the region where
$d\Gamma_1/d\rho<0$. Eventually, since these two realistic EoS are
non-causal at high densities, their relativistic fundamental
derivatives become negative.

Noteworthy, the analytic approximations of both the SLy and the FPS
EoS smoothes the jump in $\Gamma_1$ from $\simeq 1.7$ to $\simeq 2.2$
that occurs in the realistic EoSs they fit. This approximation is also
smooth across a small discontinuous drop of $\Gamma_1$ at
$\rho \simeq 2 \times 10^{14}$\,gr\,cm$^{-3}$, where muons start to
replace a part of the ultrarelativistic electrons. These non-smooth
regions of the EoS in terms of $\Gamma_1(\rho)$ may potentially yield
a negative fundamental derivative (either the classical or the
relativistic one). Our simple GGL EoS does not attempt to model the
former phenomenology.

\subsection{Limits on the values of the parameters of the GGL EoS}
\label{sec:limits}

In order to restrict the values of some of the parameters in our
phenomenological EoS it is convenient to explicitly show the
asymptotic values of the sound speed in a couple relevant regimes.

First, we consider the case in which $B=\gamma_0 -1$. Then
\begin{equation}  
\lim_{\rho \rightarrow \infty} p = (\gamma_0 - 1) \rho (1+ \varepsilon),
\end{equation}
and, therefore, the maximum sound speed allowed by the GGL EoS in this
regime is 
\begin{equation}
c_{\rm s_{(R)}}^{\rm (max)} = \sqrt{\gamma_0 -1}. 
\label{eq:csrmax}
\end{equation}
Consistently, the EoS is causal for values of $\gamma_0 $ in the
range $1 \le \gamma_0 \le 2$.

Next, we examine the limiting value of the sound speed for large values of the internal energy (i.e., the
ultrarelativistic limit). There, we may define
\begin{equation}  
\left(c_{\rm s_{(R)}}^{(+)}\right)^2  := \lim_{\varepsilon \rightarrow \infty}  c_{\rm s_{(R)}}^{2} =
\displaystyle{\gamma - 1 + \frac{d\ln{\gamma}}{d\ln{\rho}}}  = 2 ({\mathcal G}^+  - 1).
\label{eq:csR2+}
\end{equation}
For the EoS to be causal it is required that 
\begin{eqnarray}
0 \le \left(c_{\rm  s_{(R)}}^{(+)}\right)^2 \le 1, \label{eq:csr2+} \\
0 \le \left(c_{\rm s_{(R)}}^{\rm (max)}\right)^2 \le 1. \label{eq:csr2max}
\end{eqnarray}
 Hence, in order to maintain causality, it is
necessary that (from \Eqref{eq:csr2+})
\begin{equation}  
1 \le {\mathcal G}^{+} < 3/2,
\end{equation}
and (from \eqsref{eq:csrmax} and \Eqref{eq:csr2max})
\begin{equation}
1 \le \gamma_0  \le 2. 
\end{equation}


Further constraints on the values of the parameters of the GGL EoS can
be obtained demanding thermodynamic stability, which needs that
$\Gamma_1\ge 0$. This condition requires that:
\begin{eqnarray}
\gamma &\ge& 0  \label{eq:gammacondition}\\
\left(c_{\rm s_{(R)}}^{(+)}\right)^2  &\ge&  - {\displaystyle\frac{B}{\varepsilon}} \label{eq:csr2+condition}
\end{eqnarray}
Condition\,(\ref{eq:csr2+condition}) is satisfied for all non-negative values of $B$. We note that negative internal
energies are allowed by the GLL-EoS if $B>0$, with the only restriction that the pressure they produce must be positive
(see, \Eqref{eq:p(rho)}).

\subsection{Typical values}
\label{sec:typicalvalues}

Since the only constraint we have for the value of $B$ is that
$B \ge 0$ (Sec.\,\ref{sec:limits}), we will restrict ourselves to a
pair of possibilities, namely, $ B= 0$, and $B=\gamma_0 - 1$. We point
out that in the case $B=0$, if we further set $\gamma_1=\gamma_0$, the
ideal gas EoS with constant adiabatic index $\gamma=\gamma_0$ is
recovered. As we have already mentioned in \secref{sec:PhenoEoS}, the
case $B=\gamma_0 - 1$ yields an effective barotropic behavior of the
GGL EoS at high densities.

Within the ranges discussed in Sec.\,\ref{sec:limits}, we may choose
the values of the GGL EoS parameters in order to reproduce the
behaviour of, e.g., nuclear matter density. To be more specific, we
may tune the parameters of the EoS to reproduce qualitatively the
typical regimes found in the framework of core collapse supernovae
(CC-SN). With this aim, we set $\gamma_0 =4/3$ and let $\gamma_1$ vary
within the range
\begin{equation}
5/3 \le \gamma_1 \lesssim 1.9.
\label{eq:gamma1range} 
\end{equation}
Likewise, we take $\rho_1 = 1$ in dimensionless units\footnote{In this
  units, the density is scaled to some arbitrary value $\rho_{\rm sc}$
  and the pressure is measured in units of $\rho_{\rm sc}c^2$, and the
  velocity scale is $c$.}. The latter parameter is a scale factor,
whose value in CGS units, and again taking into account the studies of
CC-SN during the late stages of the infall epoch and during the bounce
epoch, $\rho_1 \lesssim 10^{15}\,$gr\,cm$^{-3}$ \citep[see,
e.g.,][]{Janka_etal:2012}.

Besides the previous analysis of the GGL EoS, we have carried out an exhaustive analysis of the EOS parameter space. We
have payed particular attention to the subspace of EoS parameters for which causality is assured for all values of
density and internal specific energy. This subspace depends strongly on the halfwidth of the gaussian sector of the
function $\gamma(\rho)$, the general trend (to preserve causality) being that we must include values of $\gamma_1$ in
the range stated above, if we set larger values of $\sigma$. Our analysis allows us to conclude that suitable values for
$\sigma$ (in units of $\rho_1$) are $0.60 \lesssim \sigma \lesssim 1.10,$ for the values of $\gamma_1$ in the range
stated in \Eqref{eq:gamma1range}.

\section{Relativistic Riemann problems}
\label{sec:RPs}

In this section we aim to set up two Riemann problems that illustrate
typical scenarios encountered in relativistic flows. In SRHD, a
Riemann problem (RP) is fully specified by providing the states on
both sides of an initial surface across which hydrodynamic variables
may exhibit discontinuities. We will refer to the uniform states on
both sides of the initial discontinuity as ``left'' and ``right''
states. Table\,\ref{tab:RiemannProblemSetup} lists the values of the
hydrodynamic variables we take as initial data for two different
RPs. The initial states have been carefully chosen to serve as
prototypes of different dynamical situations, some of which are
genuinely relativistic (i.e., they cannot be found in classical
hydrodynamics), inasmuch as some of the initial states fall in a
thermodynamic region of non-convexity from the relativistic point of
view, but not from the classical point of view (see below).
\begin{table*}
\centering
\caption{Set up of the three Riemann problems considered in Sec.\,\ref{sec:RPs}. The first column provides a short
  text-tag to identify the problem set up. Columns 2-4 (5-7) display the values of the rest-mass density, velocity and
  pressure left (right) to the initial discontinuity.}
\label{tab:RiemannProblemSetup}
\begin{tabular}{lcllllll}
\hline
{\text Riemann Problem} & Acronym & $\rho_L$ &  $v_L$ &  $p_L$ &  $\rho_R$ &  $v_R$ &  $p_R$  \\\hline
{\text Blast Wave}  & BW & $1.0$ &  $0$ &  $1000$ &  $0.125$ &  $0$ &  $0.1$  \\
{\text Colliding Slabs}  & CS&  $0.125$ &  $0.99$ &  $1.0$ &  $0.125$ &  $-0.99$ &  $1.0$  \\
\hline
\end{tabular}
\end{table*}
\begin{figure*}
\centering{
\includegraphics[width=0.48\textwidth]{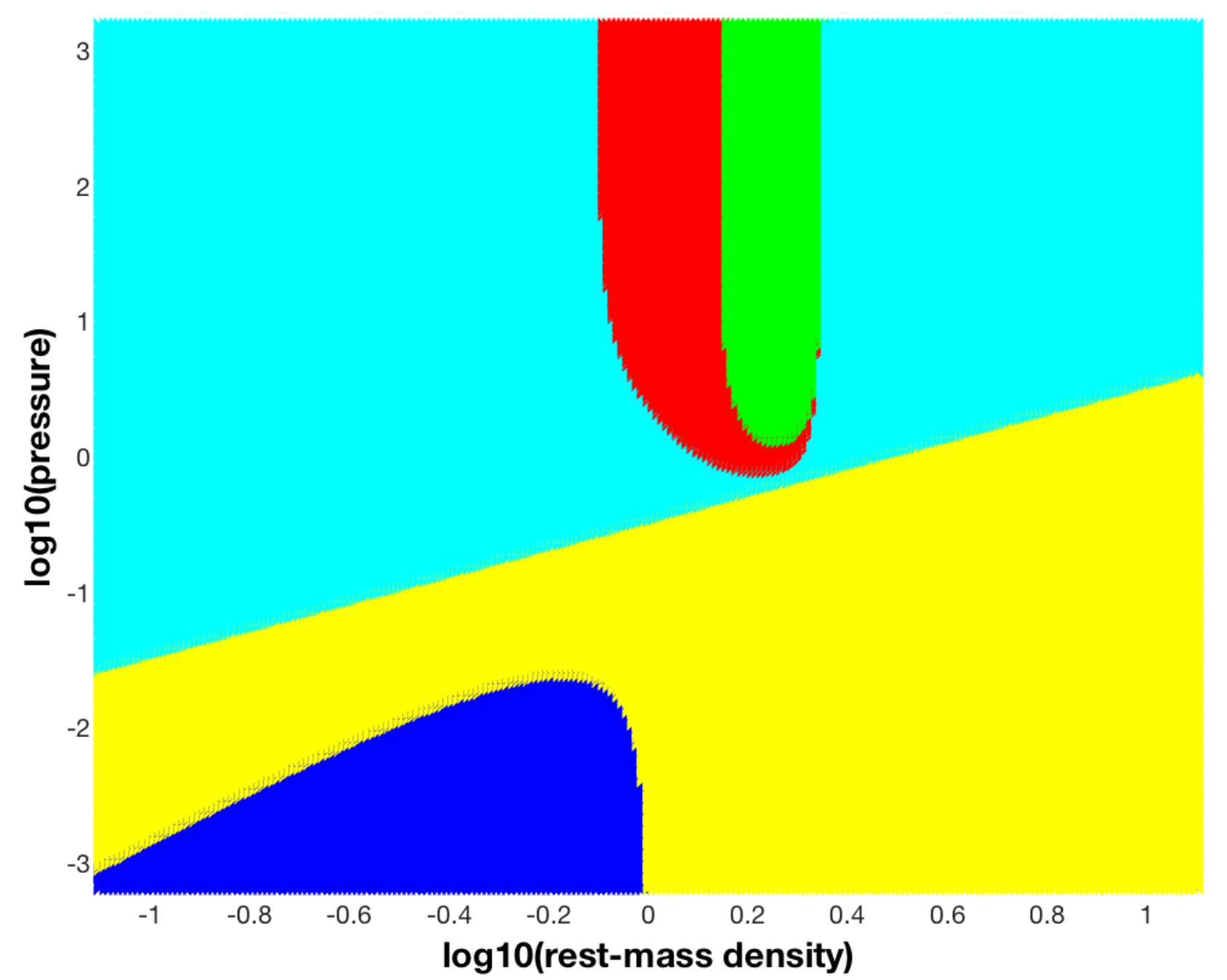} 
\includegraphics[width=0.48\textwidth]{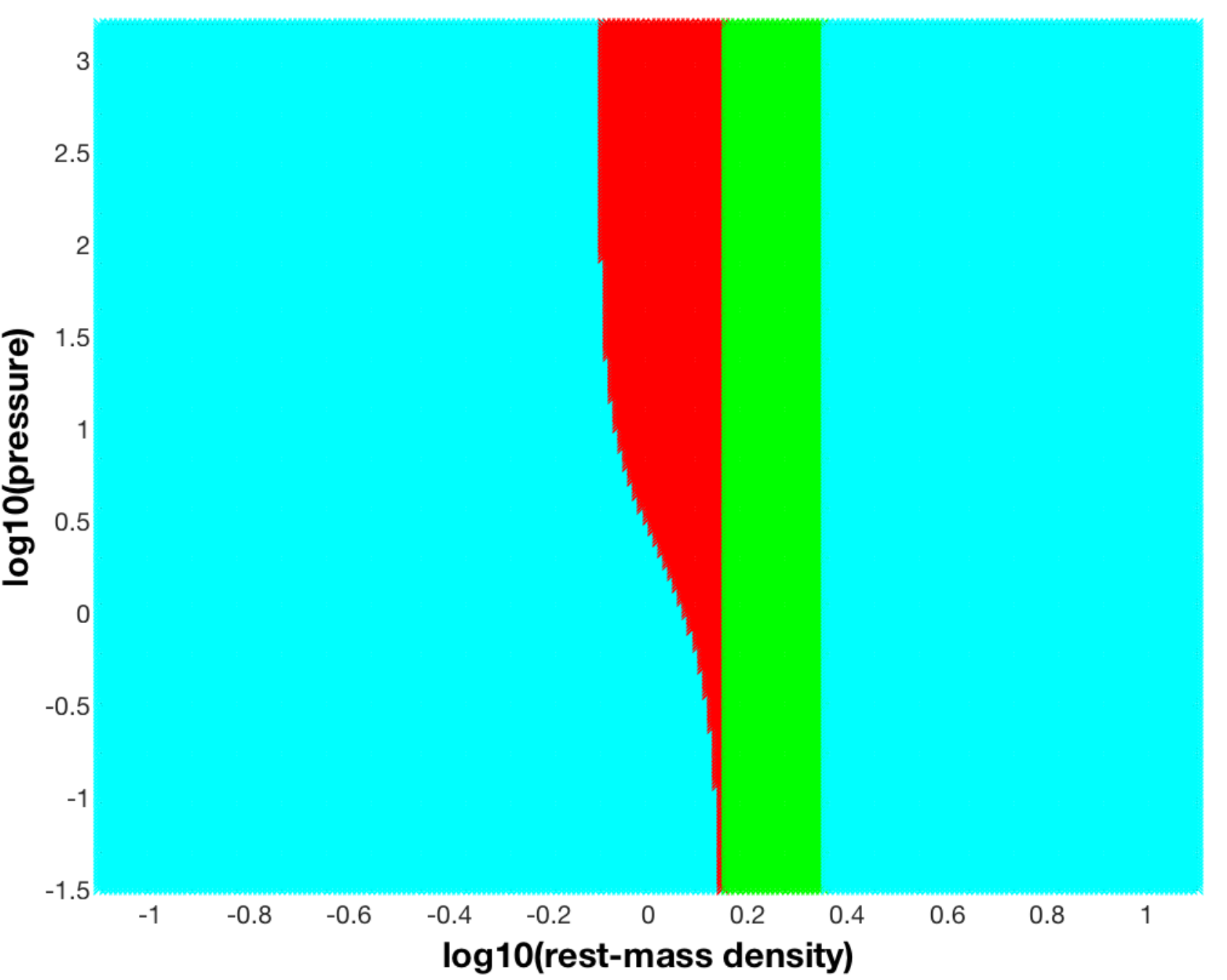}}
\caption{$p-\rho$ diagrams in logarithmic scale of the GGL EoSs
  defined with parameters in of models EoS1 (left) and
    EoS2 (right) (see \tabref{tab:EoSparameters}).}
\label{fig:diagrams}
\end{figure*}

In order to understand the nonclassical behavior of the Riemann fan in
relativistic hydrodynamics closed with a non-convex EoS, we present
two $p-\rho$ diagrams in logarithmic scale of the GGL EoS
(\figref{fig:diagrams}) defined with parameters in
\tabref{tab:EoSparameters}. We note that with the parameter set EoS3,
the GGL EoS reduces to an ideal gas EoS with constant adiabatic index
$\gamma=4/3$ which, therefore, does not have any non-convex region. In
the following, we will outline the most salient features of the two
RPs defined by their initial states (\tabref{tab:RiemannProblemSetup})
employing the two parameterizations of the GGL EoS that encompass
non-convex regions in the $p-\rho$ (EoS1 and EoS2). We will further
compare the non-standard wave pattern that may develop with the latter
parameterizations of the EoS with the classical wave patterns that
arise when using the EoS3 parameter set.
\begin{table}
\centering
\caption{Parameters of the GGL EoS for two prototype thermodynamic situations (EoS1 and EoS2), as well as for a pure
  ideal gas EoS with constant adiabatic index (EoS3).}
\label{tab:EoSparameters}
\begin{tabular}{lllll}
\hline
Model & $\gamma_0$ &  $\gamma_1$ &  $\sigma$ &  $B$  \\
\hline
EoS1  & $4/3$ &  $1.9$ &  $1.1$ &  $\gamma_0-1$   \\
EoS2 & $4/3$ &  $1.9$ &  $1.1$ &  $0$  \\
EoS3 & $4/3$ &  $4/3$ &  $1.0$ & $0$  \\\hline
\end{tabular}
\end{table}

The diagram on the left in \figref{fig:diagrams} corresponds to the
$p-\rho$ plane of the GGL EoS employing the parameters of model EoS1
(\tabref{tab:EoSparameters}). Different from the EoS2 and EoS3 models,
the EoS1 parameter set features a value $B=\gamma_0-1$, which leads to
a high-density EoS of barotropic type. We observe five different
regions represented with distinct colors.  The area in cyan color
depicts the region where both ${\mathcal G}_{\rm (C)}>0$ and
${\mathcal G}_{\rm (R)}>0$.  In this region the EoS is convex, both
classically and from the relativistic point of view. Hence, a
classical wave structure is expected in the Taub adiabat connecting
the left and right states. The area in green represents the region
where ${\mathcal G}_{\rm (C)}<0$ and, consistently,
${\mathcal G}_{\rm (R)}<0$. This is a region of non-convexity both in
the classical and in the relativistic sense. The section in red
corresponds to the case where ${\mathcal G}_{\rm (C)}>0$ and
${\mathcal G}_{\rm (R)}<0$, which defines a purely relativistic
non-convex region. The sector in yellow color responds to the region
where the specific internal energy is negative. This region of the
parameter space may also arise in realistic nuclear matter EoSs as a
result of the (negative) contribution of the interaction potential
between different matter constituents, but is typically restricted to
a small region around nuclear saturation density
\citep[e.g.,][]{Engvik:1996ApJ...469..794E,Mansour:2012PAN....75..430M}.
Finally, the small dark blue area, surrounded by the yellow one, shows
the pathological region where the SRHD equations loose hyperbolicity
as $c_{\rm s(C)}^2<0$.

\begin{figure}
\centering
\includegraphics[width=0.48\textwidth]{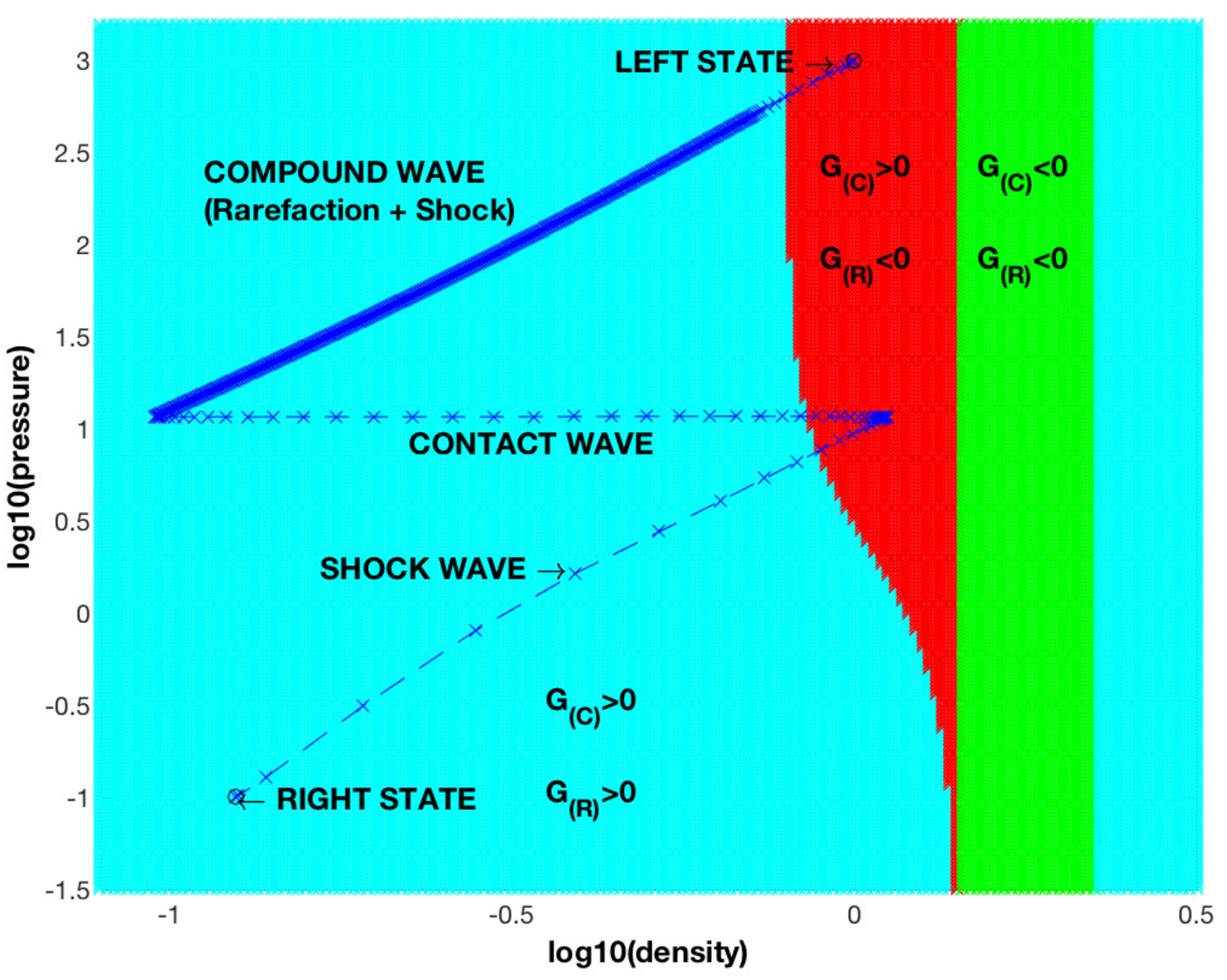}
\caption{Numerical approximation of the locus of the gas states
  connecting the left and right states in the $p-\rho$ (EoS2)
  plane for the BW Riemann problem}
\label{fig:BWdiagramaEOS2}
\end{figure}

\begin{figure*}
\centering{
\includegraphics[width=0.48\textwidth]{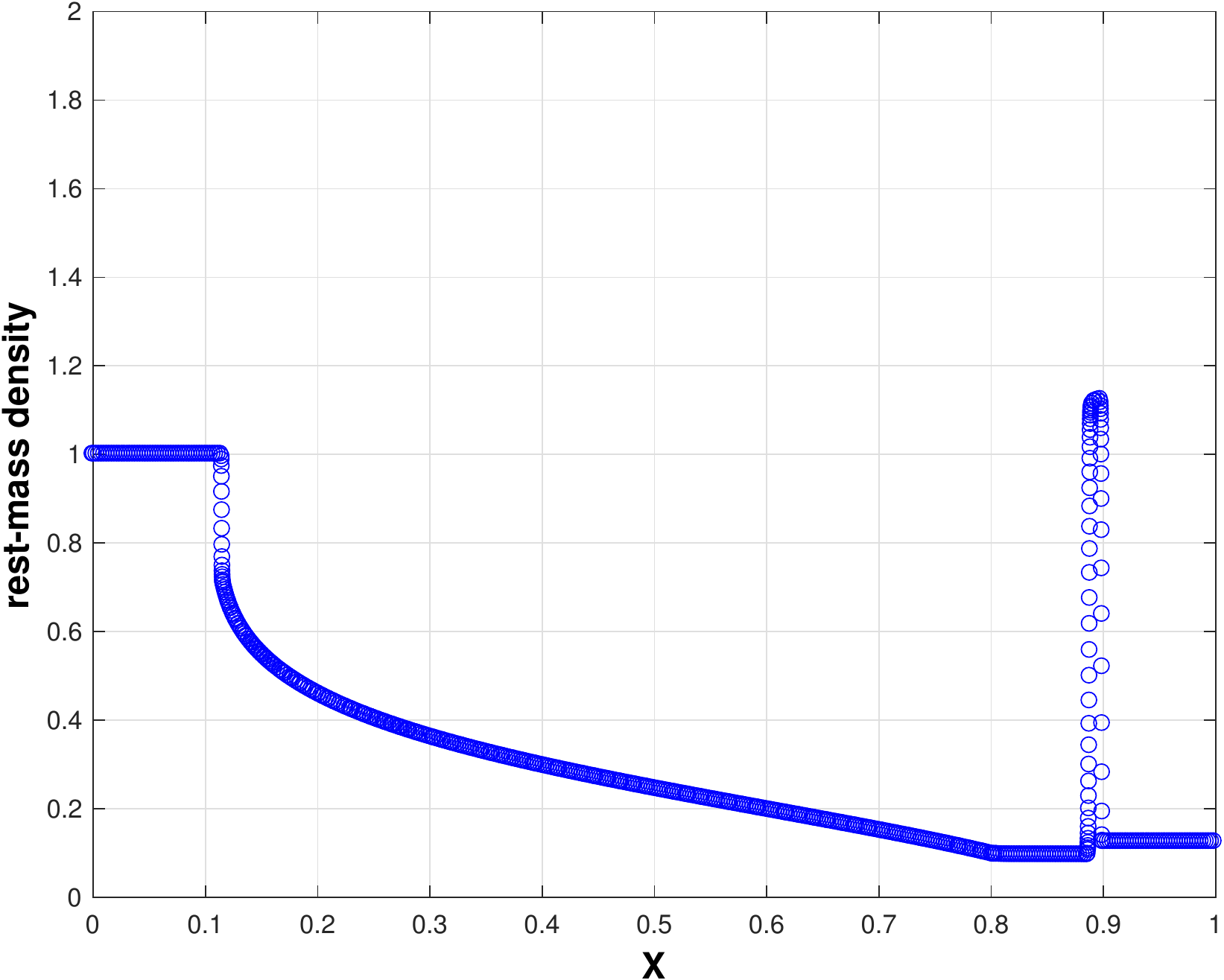}\hskip 12pt
\includegraphics[width=0.48\textwidth]{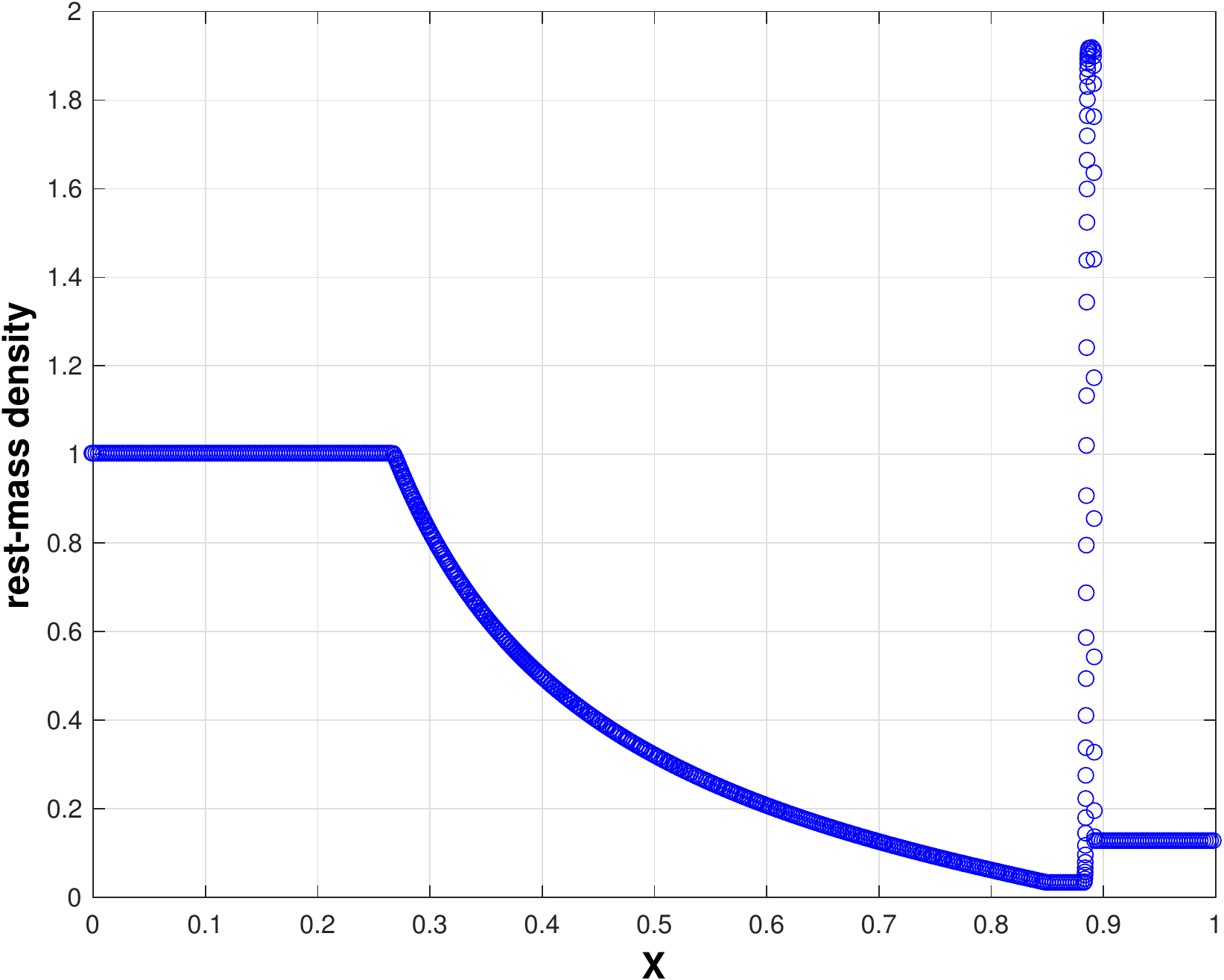}}\\
\vskip 8pt
\centering{
\includegraphics[width=0.48\textwidth]{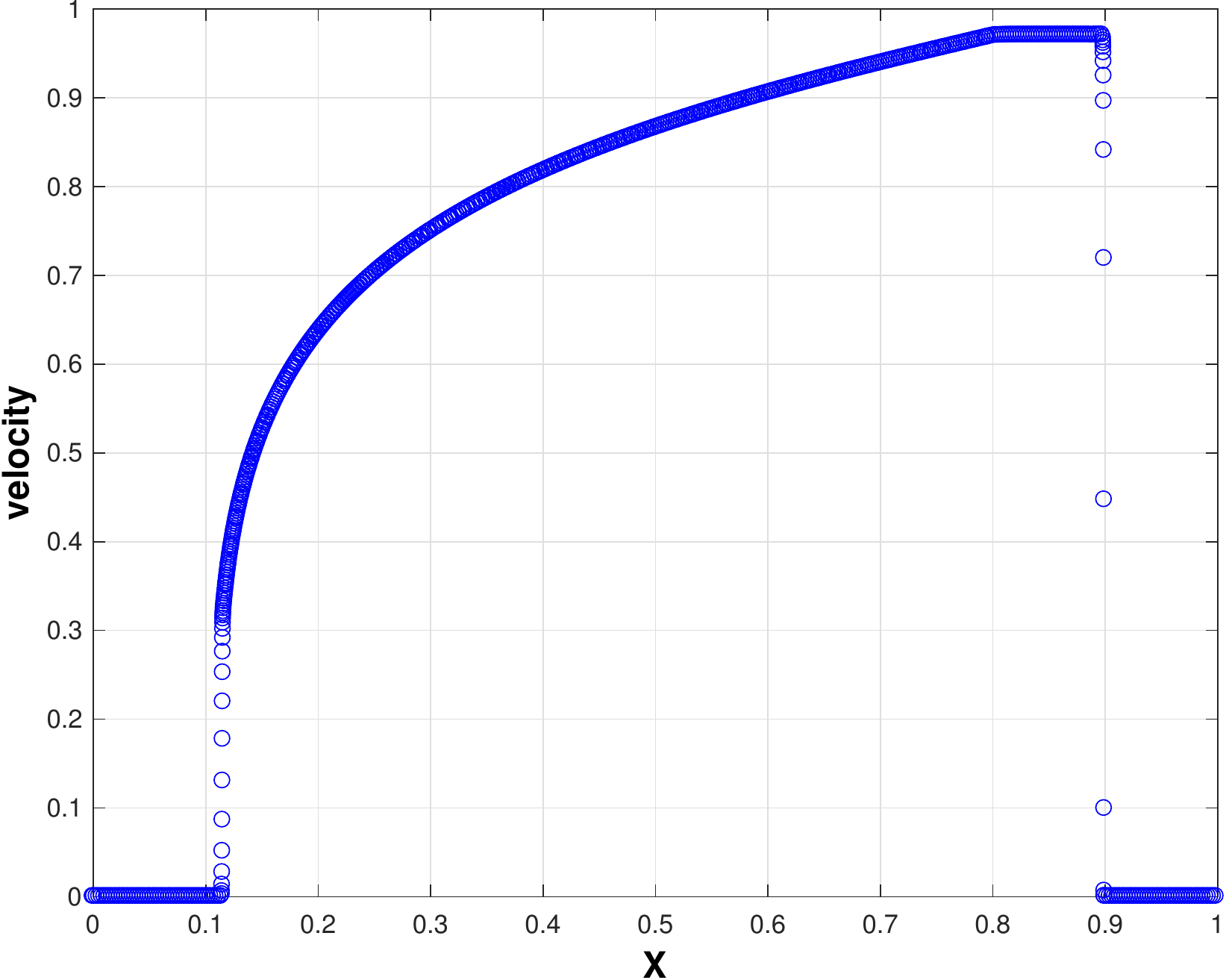}\hskip 12pt
\includegraphics[width=0.48\textwidth]{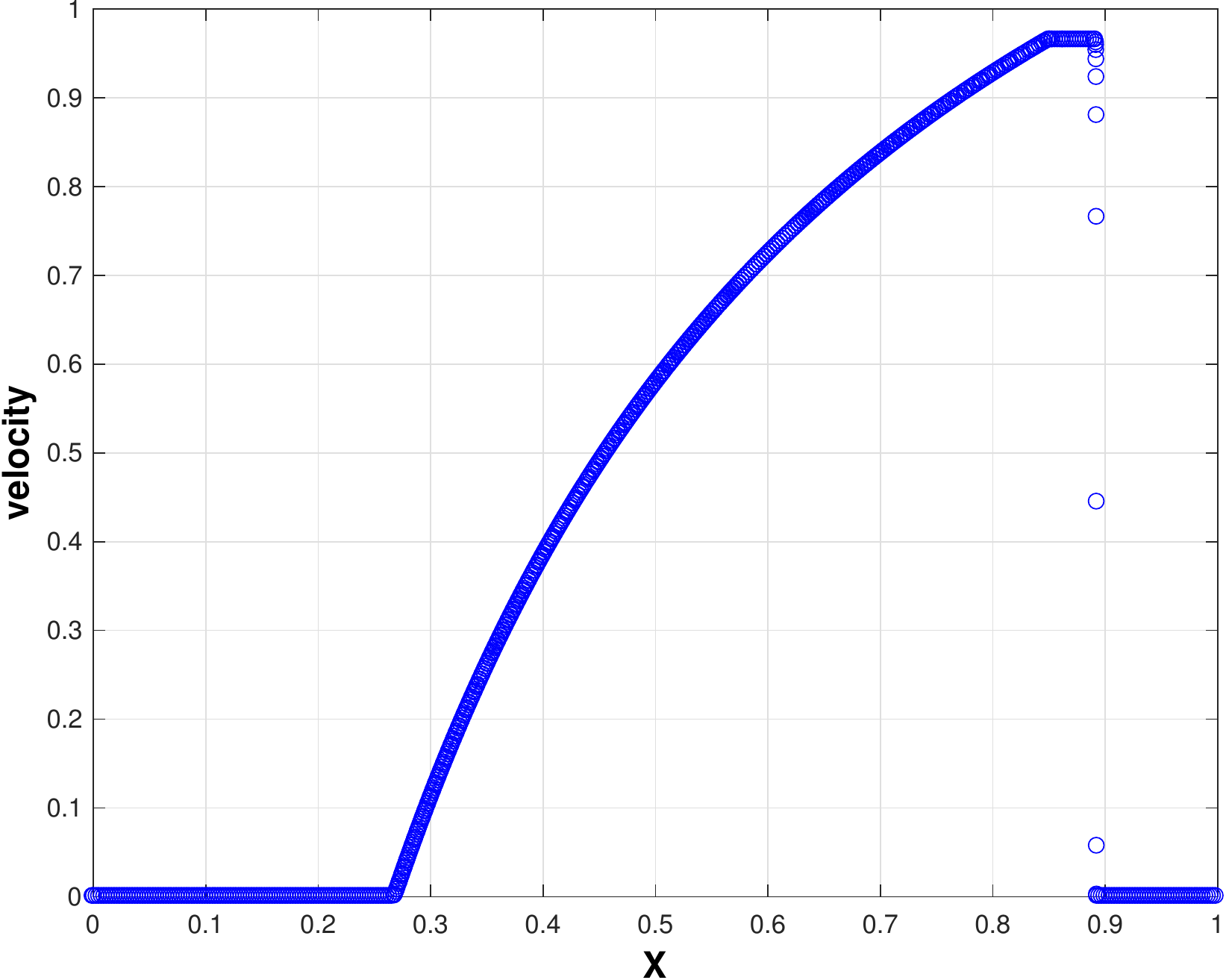}}\\
\vskip 8pt
\centering{
\includegraphics[width=0.48\textwidth]{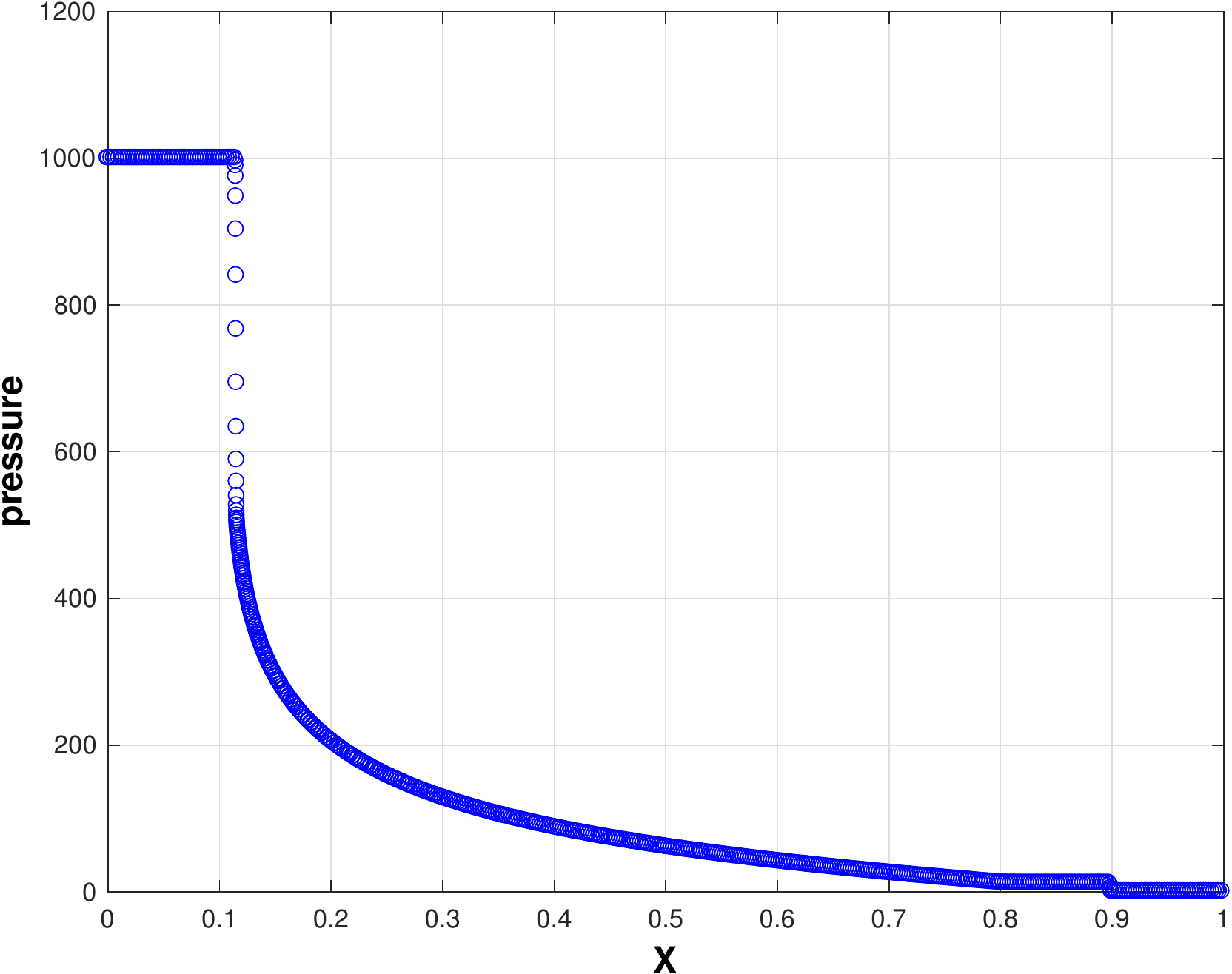}\hskip 12pt
\includegraphics[width=0.48\textwidth]{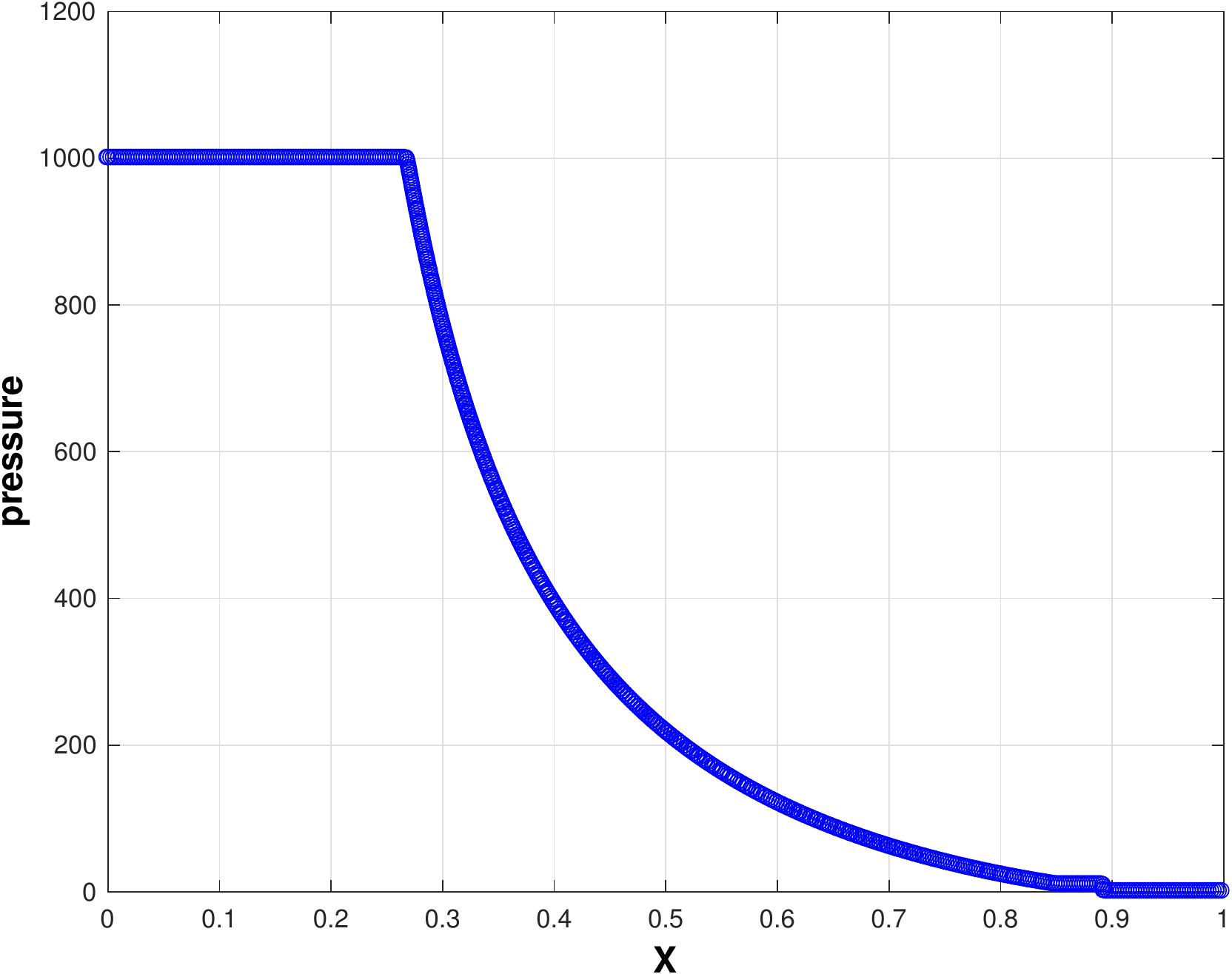}}
\caption{Left column: Rest-mass density (upper left), velocity
  (middle) and pressure (bottom) of the relativistic blast wave RP set
  according to \tabref{tab:RiemannProblemSetup} with the GGL EoS and
  the parameters corresponding to the EoS2 model
  (\tabref{tab:EoSparameters}). Right column: same as left column but
  for the relativistic blast wave RP set according to
  \tabref{tab:RiemannProblemSetup} with the ideal gas EoS
  corresponding to the EoS3 model.}
\label{fig:BWrelativistic}
\end{figure*}

\begin{figure}
\centering
\includegraphics[width=0.48\textwidth]{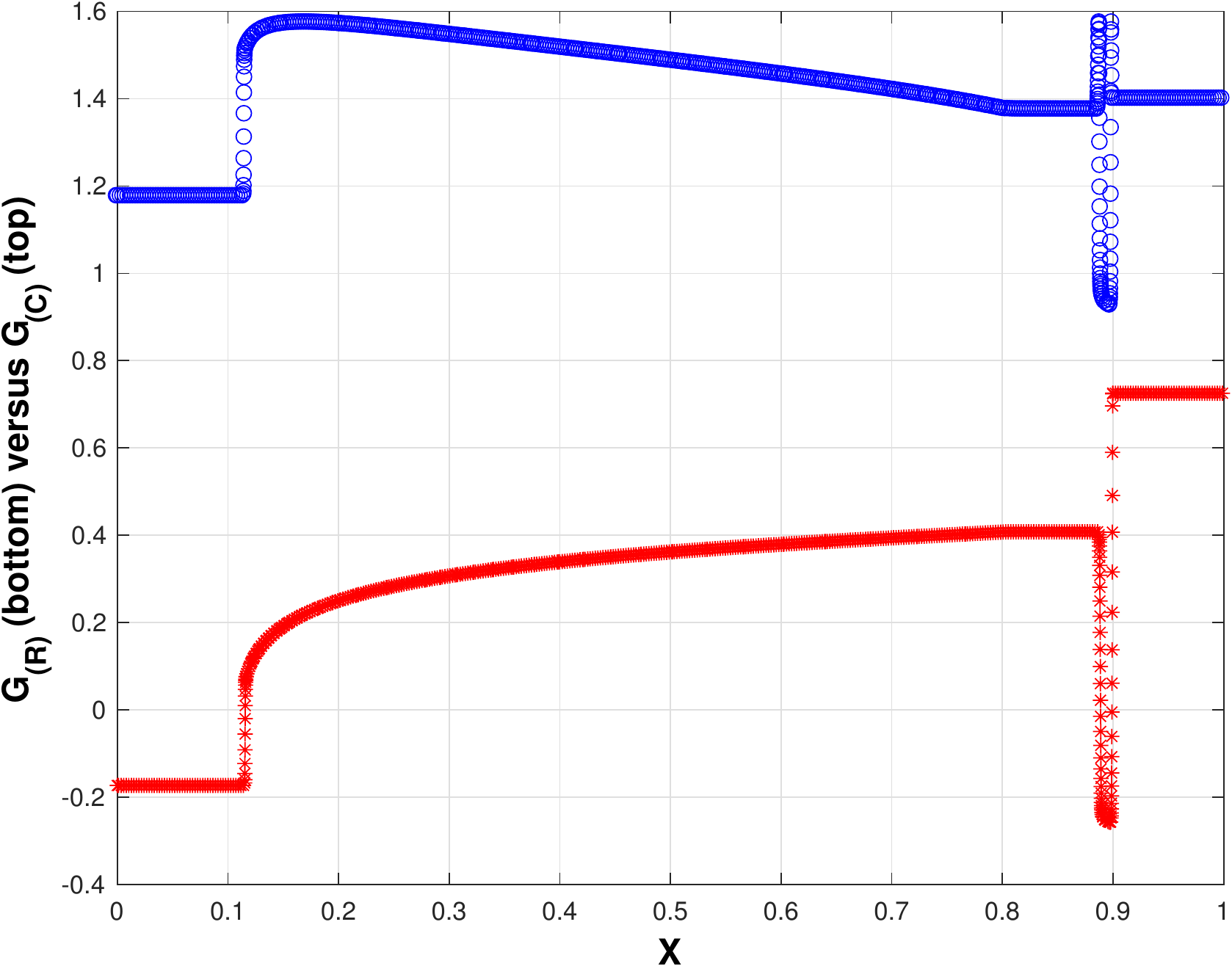}
\caption{Classical (blue circles) versus relativistic (red asterisks)
  fundamental derivative profiles of the relativistic blast wave RP
  set according to \tabref{tab:RiemannProblemSetup} with the GGL EoS
  and the parameters corresponding to the EoS2 model
  (\tabref{tab:EoSparameters}).}
\label{fig:BW_FundamentalD}
\end{figure}

\begin{figure*}
\centering{
\includegraphics[width=0.48\textwidth]{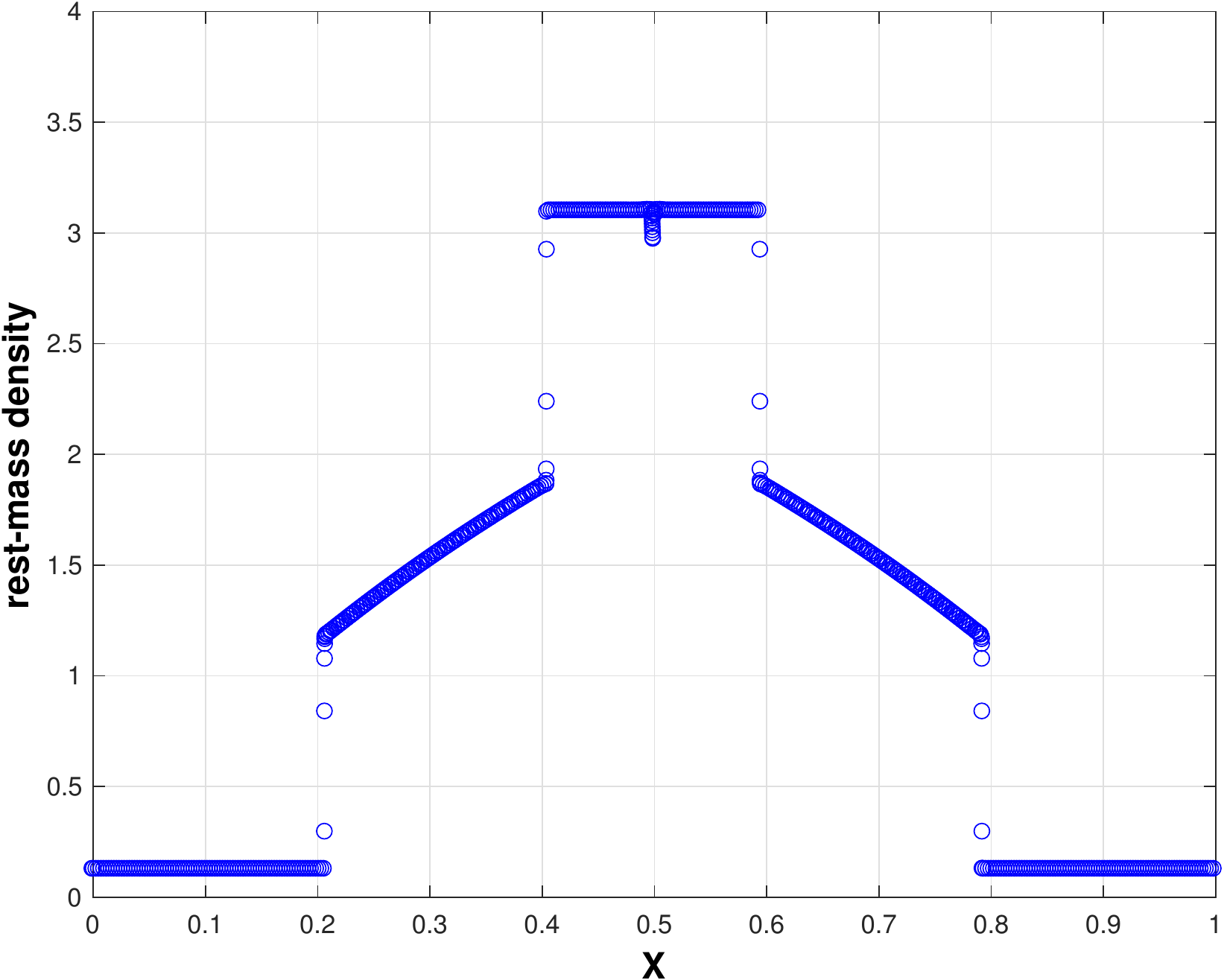}\hskip 12pt
\includegraphics[width=0.48\textwidth]{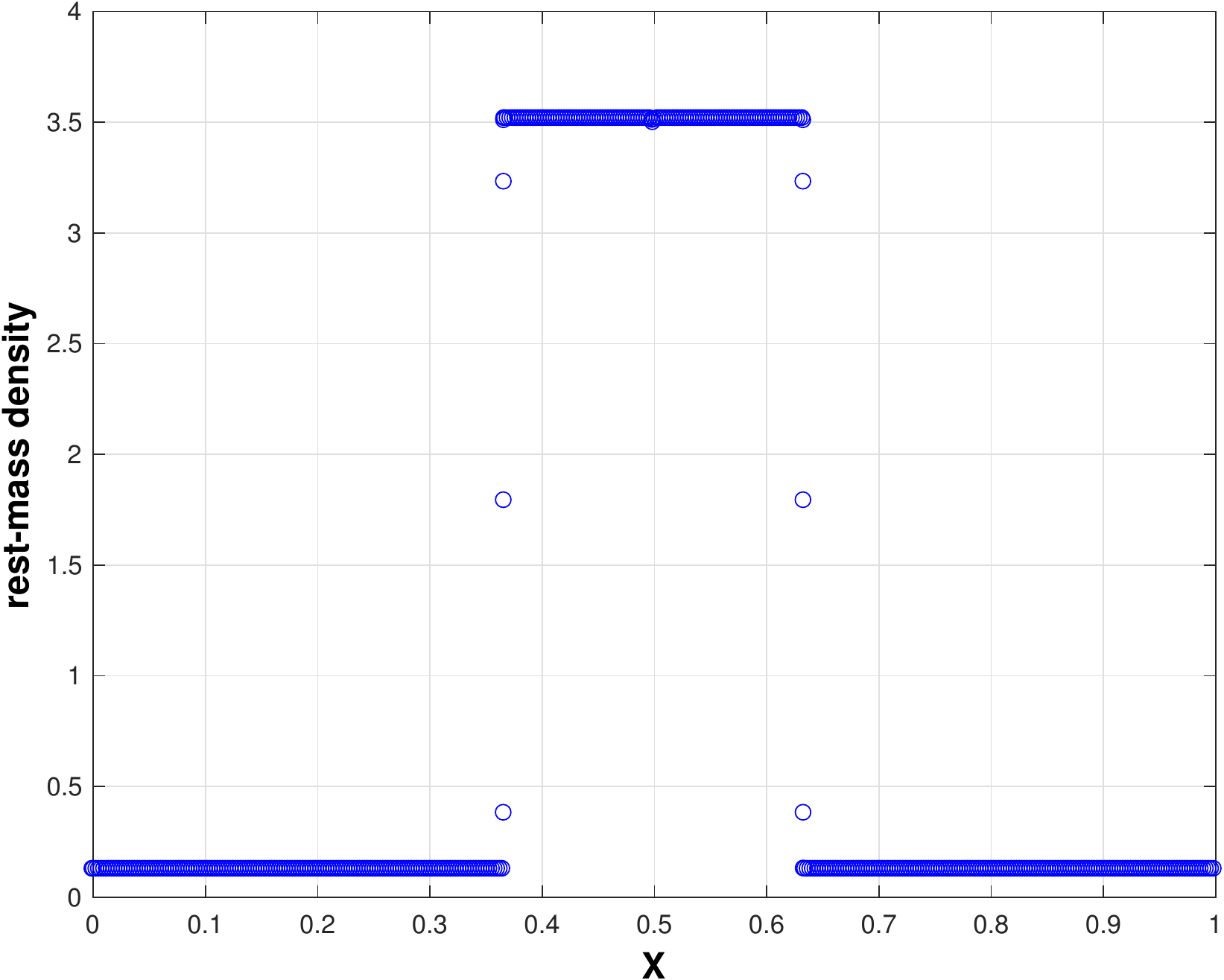}}\\
\vskip 8pt
\centering{
\includegraphics[width=0.48\textwidth]{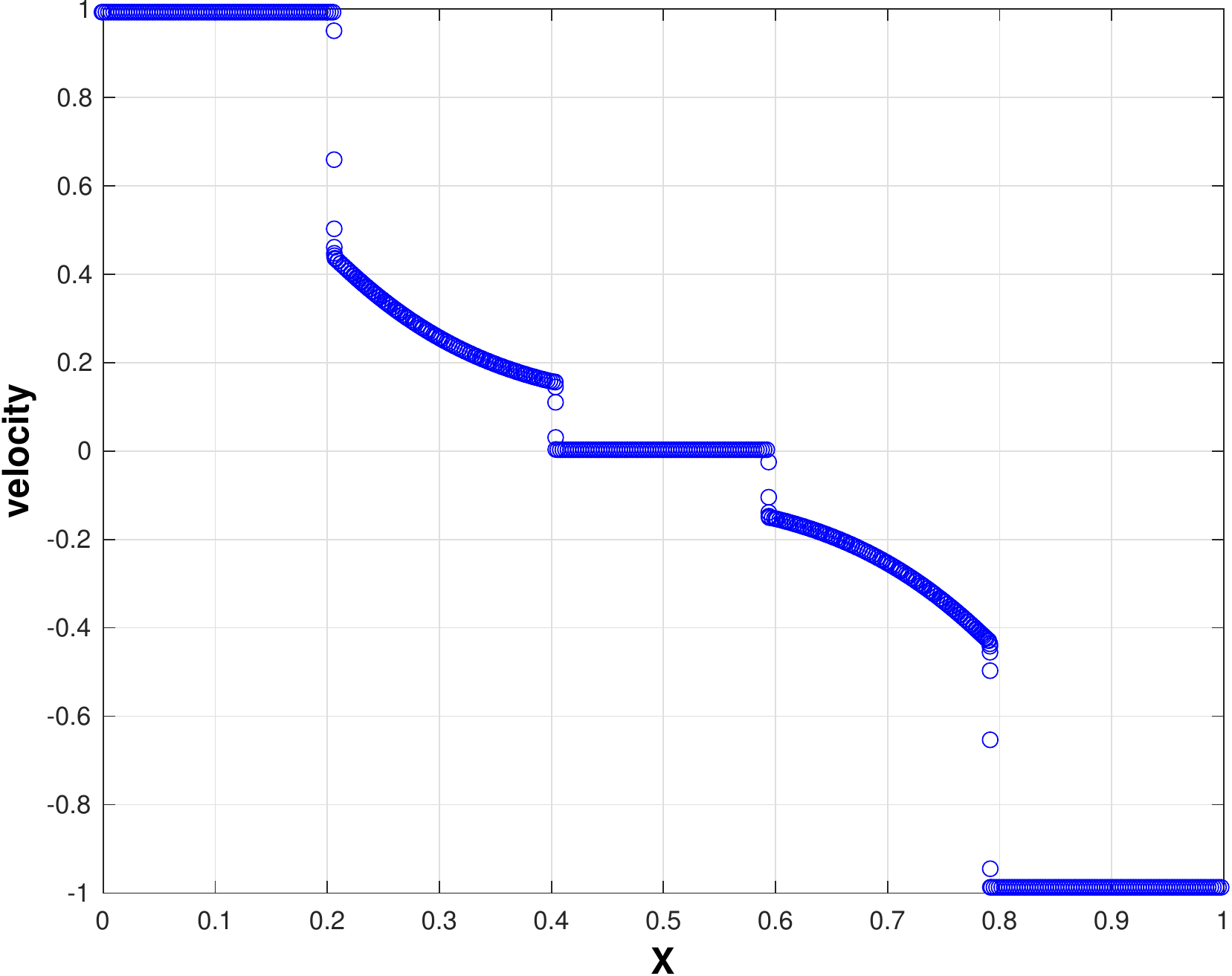}\hskip 12pt
\includegraphics[width=0.48\textwidth]{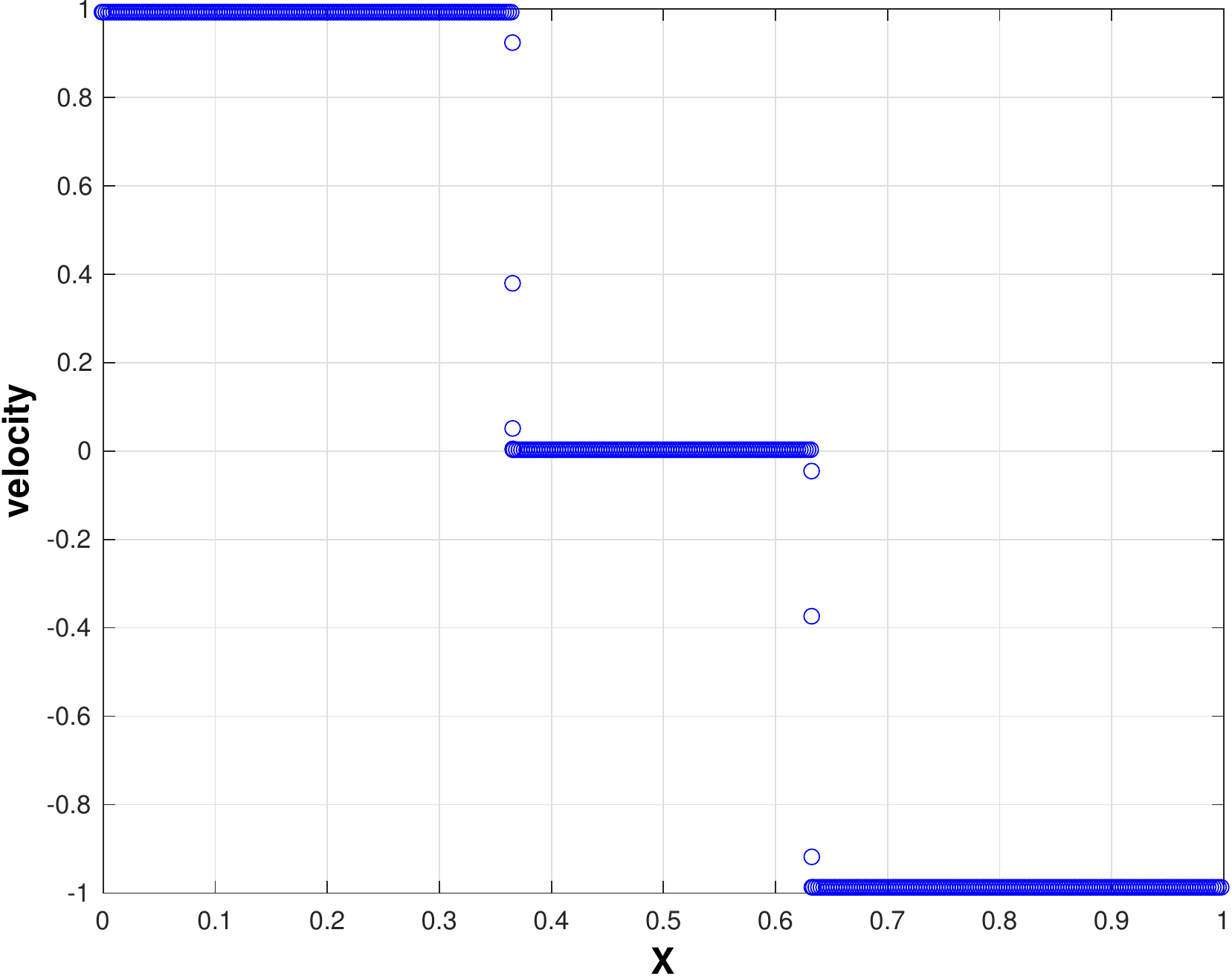}}\\
\vskip 8pt
\centering{
\includegraphics[width=0.48\textwidth]{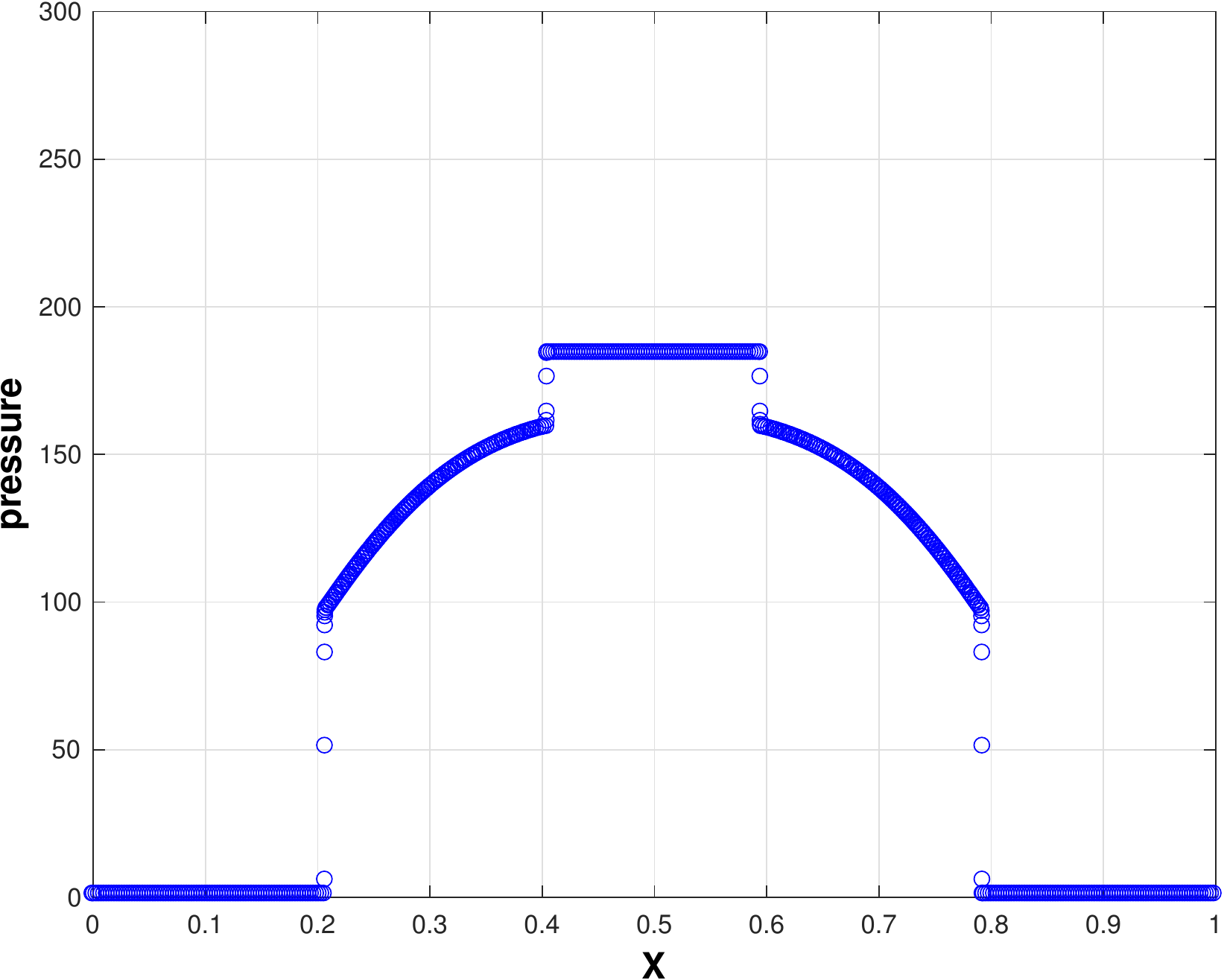}\hskip 12pt
\includegraphics[width=0.48\textwidth]{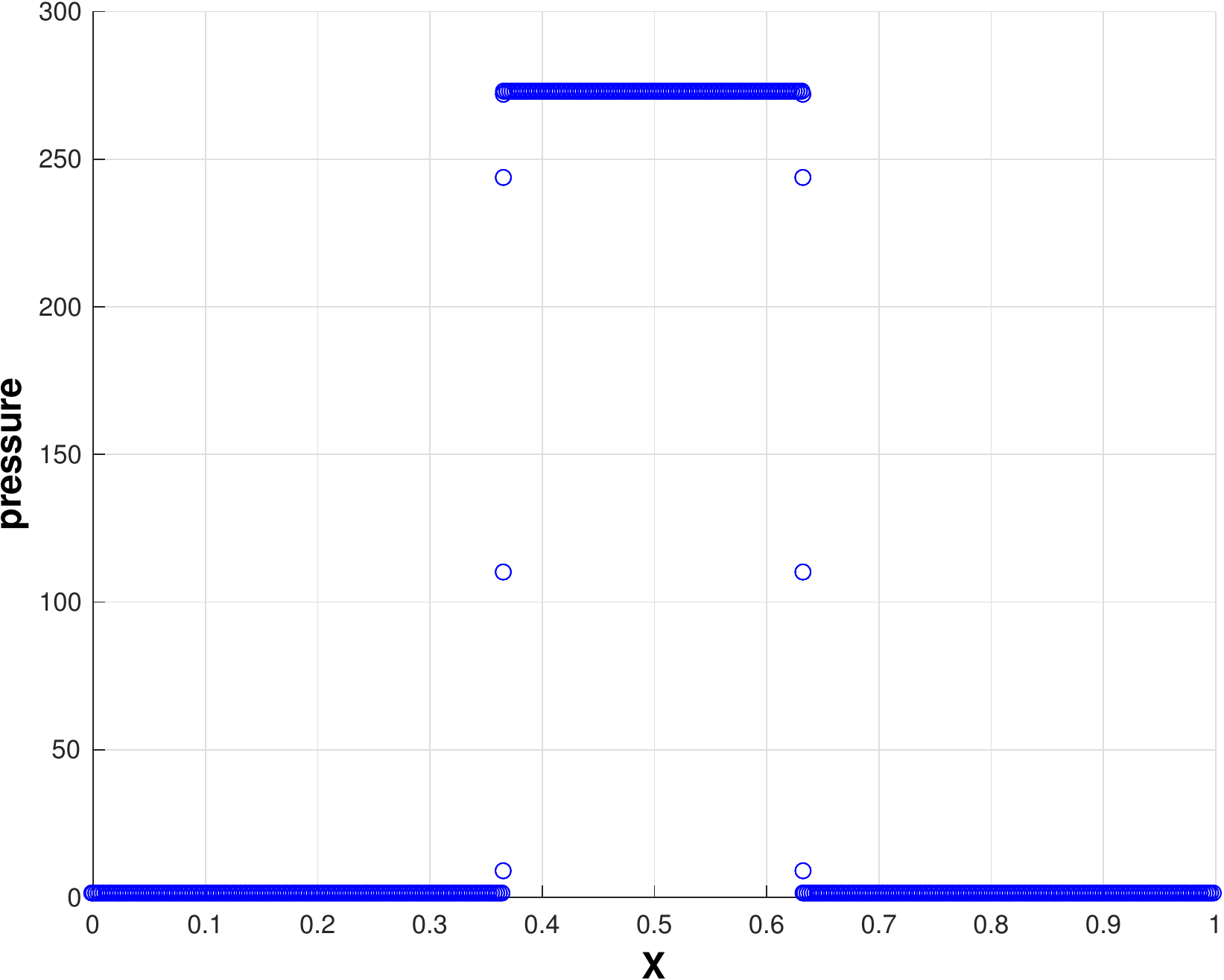}}
\caption{Left column: Rest-mass density (upper left), velocity
  (middle) and pressure (bottom) of the relativistic colliding slabs
  RP set according to \tabref{tab:RiemannProblemSetup} with the GGL
  EoS and the parameters corresponding to the EoS1 model
  (\tabref{tab:EoSparameters}). Right column: same as left column but
  for the relativistic colliding slabs RP set according to
  \tabref{tab:RiemannProblemSetup} with the ideal gas EoS
  corresponding to the EoS3 model.}
\label{fig:CSlabs}
\end{figure*}

\begin{figure}
\centering
\includegraphics[width=0.48\textwidth]{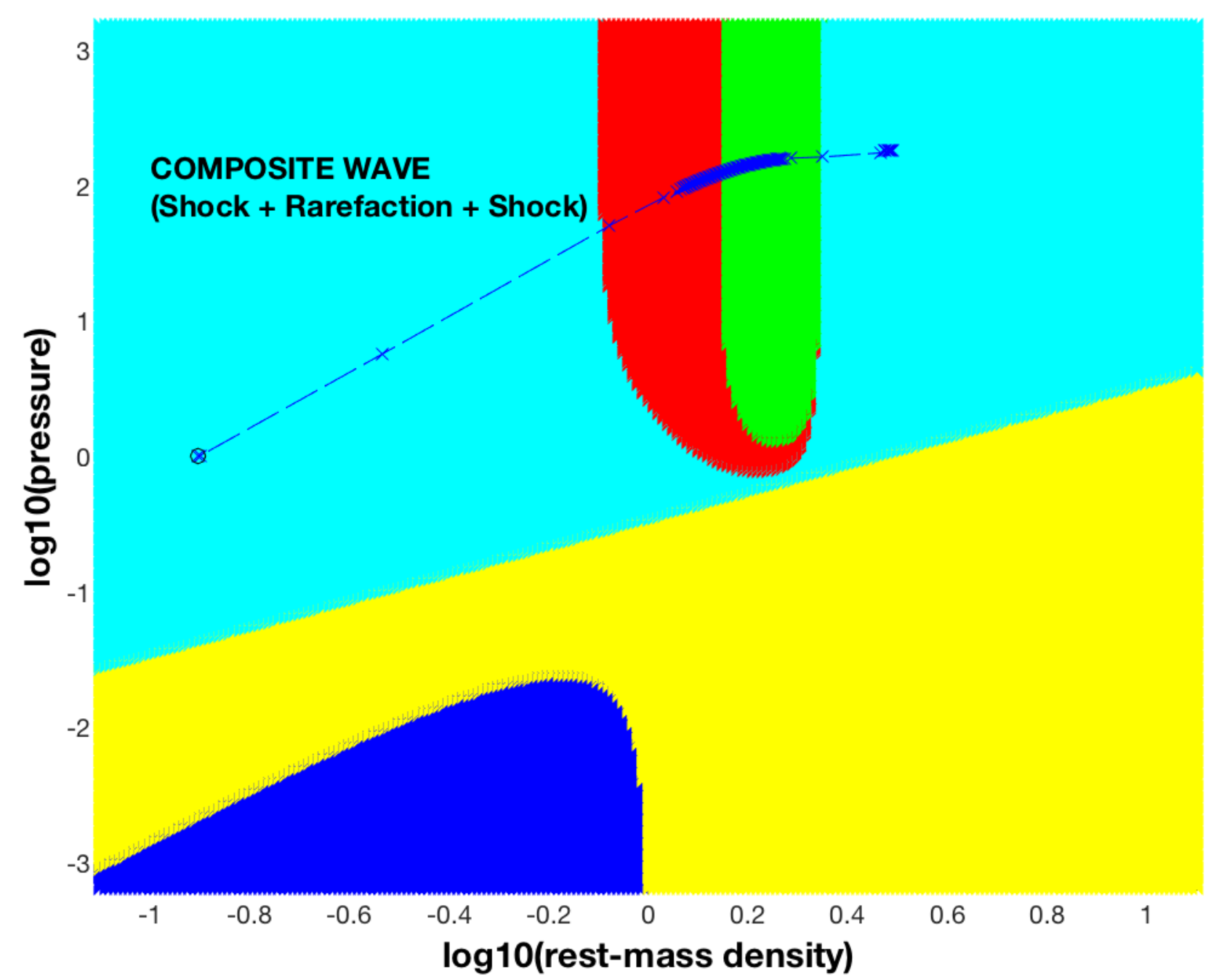}
\caption{Numerical approximation of the locus of the gas states
  connecting the left and right states in the $p-\rho$ plane
  (EoS1) for the CS Riemann problem}
\label{fig:CSdiagramaEOS1_Bneq0}
\end{figure}

\begin{figure}
\centering
\includegraphics[width=0.48\textwidth]{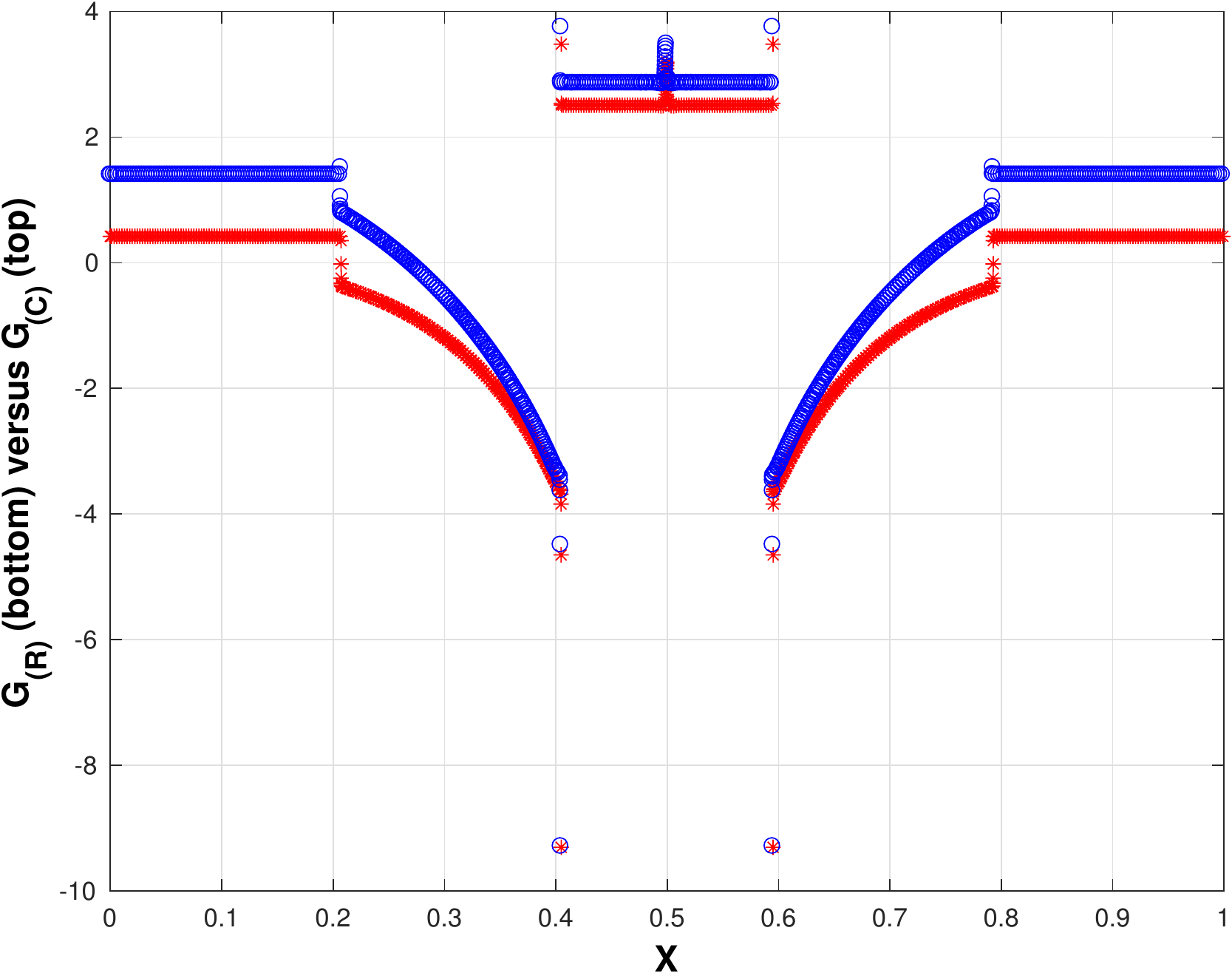} 
\caption{Classical (blue circles) versus relativistic (red asterisks)
  fundamental derivative profiles of the relativistic colliding slabs
  RP set according to \tabref{tab:RiemannProblemSetup} with the GGL
  EoS and the parameters corresponding to the EoS1 model
  (\tabref{tab:EoSparameters}).}
\label{fig:CS_FundamentalD}
\end{figure}

The analogous regions for the GGL EoS setup EoS2 (\tabref{tab:EoSparameters}) are represented in the diagram on the
right of \figref{fig:diagrams}.  The different colors designate same features as in the diagram on the left. If $B=0$ 
it is straightforward to prove that ${\mathcal G}_{\rm (C)}$ only depends on the density. This feature explains the regular shape of the green vertical band.  Another consequence is that both regions of loose of hyperbolicity (dark blue) and negative value of
specific internal energy (yellow) are avoided.

We perform numerical simulations by means of a local characteristic approach  based on the Marquina Flux Formula for Special Relativistic flows \citep{Martietal:1997ApJ}. In order to resolve the complex wave structure we use a similar strategy as followed in \cite{Serna:2009JCP} and \cite {Sernaetal:2014PoF} to capture composite waves in magnetohydrodynamics.
The numerical scheme uses two linearizations at each side of the interface  computing the numerical fluxes  through a Lax-Friedrichs approach ensuring stability and entropy satisfying solution. We  implement high order accuracy in space following the Shu-Osher flux formulation \citep{ShuOsher:1989JCP} by using a third order reconstruction procedure based on hyperbolas \citep{Marquina:1994SISC}.  High order accuracy in time is achieved by a third order TVD Runge-Kutta time stepping procedure \citep{ShuOsher:1989JCP}. 

Unless stated otherwise, in the following, we will show the solution
of the RPs at a time $t=0.4$ after the initial discontinuity begins to
break up ($t=0$). The exact value of the final time to represent the
solution is physically irrelevant since the solution of the RP is
self-similar in SRHD and planar geometry. The value we take here is
adequate to show the details of the Riemann wave pattern in various
figures of this section. The solution is computed with $12800$ uniform
grid points and a CFL factor $\CFL=0.4$.

We first compare the evolution of a blast wave with the GLL-EoS and
parameters of the sets EoS2 and EoS3 (\tabref{tab:EoSparameters}). The
initial data in \tabref{tab:RiemannProblemSetup} (BW) corresponds to a
relativistic strong blast wave such that the left state belongs to the
red region (the purely relativistic non-convex region) in the diagram
on the right in \figref{fig:diagrams}.  The right state belongs to the
cyan region, (the pure classical convex
region). \figref{fig:BWdiagramaEOS2} shows the numerical approximation
of the path connecting the left and right states in the $p-\rho$ plane
for the RP at hand.  The profiles of the wave structure in rest-mass
density, velocity and pressure are shown in the left panels of
\figref{fig:BWrelativistic}. These exhibit a path connecting both
states, consisting of a shock wave propagating to the right and a
rarefaction wave attached to a shock wave (composite wave), separated
by a contact wave. Figure\,\ref{fig:BW_FundamentalD} shows the
profiles of the classical and relativistic fundamental
derivatives. The latter shows that the left wave breaks into two
pieces at the point where the relativistic fundamental derivative
changes sign, forming a composite wave (shock plus rarefaction). We
note that should we chose the left state in the classical non-convex
region (green region in \figref{fig:BWdiagramaEOS2}), the resulting
wave pattern would have been qualitative similar. However, this RP
example is genuinely relativistic and shows that beyond the classical
region of convexity loss, relativistic effects my drive a non-convex
dynamics.


For comparison, we provide in the right panels of
\figref{fig:BWrelativistic} the evolution of the same initial data as
in the BW case (\tabref{tab:RiemannProblemSetup}), but for an ideal
gas EoS with $\gamma=4/3$, corresponding to the tag EoS3 in
\tabref{tab:EoSparameters}.  The profiles of the wave structure in
density, velocity and pressure show the classical wave structure of a
relativistic blast wave, an expansive rarefaction starting at the left
state, a contact and a shock wave. 

We have not included a variant of the relativistic blast wave problem
with the EoS1 parameter set. The reason is that the Taub adiabat
joining the initial states for this test
(\tabref{tab:RiemannProblemSetup}) does not cross over any non-convex
region. Hence, the resulting wave pattern is qualitatively equal to
the one generated with the EoS3 parameter set of the GGL EoS.

The left panels of \figref{fig:CSlabs} exhibit the evolution at time $t=0.4$ of two relativistic colliding slabs with
GGL EoS parameters of the set EoS1.  The initial data in \tabref{tab:RiemannProblemSetup} consist of two streams of gas
hitting each other with opposite and equal speeds of $0.99$. Both the density and the pressure of the left and right
states of the RP are uniform and identical. Their values have been chosen in such a way that the left and right states
belong to the cyan region of \figref{fig:diagrams} (right panel) and are located at the same point. Therefore, both
states fall in a region where the EoS is convex, but as we observe, the Taub adiabat joining them transits over the
non-convex realm of the EoS, as shown in the diagram in    \figref{fig:CSdiagramaEOS1_Bneq0},  resulting into a striking dynamics. The solution is symmetric with respect to the center of
the domain because the initial velocities are prescribed with the same magnitude and opposite sign in order to hit the
two gases at the center.  The wave structure consists of two composite waves traveling from the center in opposite
directions.  Each composite wave consist of three pieces, two shock waves connected through a rarefaction wave.  The
profile of the classical and relativistic fundamental derivatives in \figref{fig:CS_FundamentalD} reveals
the region of negative values where the anomalous wave structure appears.
Employing an ideal EoS (EoS3 in \tabref{tab:EoSparameters}) the
initial discontinuity breaks up into just two symmetric shock waves
traveling towards the boundary of the domain at constant velocity
(right panels of \figref{fig:CSlabs}). Should we use the EoS2
parameter set for the GGL EoS, we would have obtained a qualitatively
equal wave pattern than obtained with the EoS3 parameter set, since
the Taub adiabat does not pierce the non-convex regions of the
$p-\rho$-plane for the initial states of the relativistic colliding
slabs RP.

\section{Discussion and conclusions}
 \label{sec:summary}

 The aim of this paper is to show the rich and complex dynamics that
 an EoS with non-convex regions in the $p-\rho$ plane may develop as a
 result of genuinely relativistic effects, without a classical
 counterpart. To best serve our purposes, we have introduced a
 phenomenological EoS, the GLL-EoS, which depends on five parameters
 that can be restricted heeding to causality and thermodynamic
 stability constraints. This EoS shall be regarded as a toy-model with
 which we may mimic realistic (and far more complex) EoS of practical
 use in the realm of Relativistic Hydrodynamics. Actually, causality
 and thermodynamic stability restrictions provide an ample playground
 for the variation of the GLL-EoS parameters. We exploit this freedom
 to calibrate the parameters so that our simple closure reproduces
 some qualitative behaviors found in microphysical EoSs of matter
 around and above nuclear saturation density. In particular, the
 non-monotonic dependence of the sound speed (equivalently of the
 adiabatic index) with the rest-mass density. Certainly, the
 non-monotonic dependence of $\Gamma_1$ on $\rho$ can also be found in
 many other contexts within Astrophysics (e.g., in problems involving
 partially ionized plasma) and elsewhere (e.g., in material
 processing), and it may not always result in a non-convex
 thermodynamics. The latter depends on whether this lack of
 monotonicity translates into a negative fundamental derivative.

 The dynamic effects of a non-convex EoS gleam when discontinuities
 are set in the fluid or result from its non-linear
 evolution. Studying the breakup of an initial discontinuity in the
 flow is a simple yet effective way of exploring the aforementioned
 dynamics. Fortunately, there exist a physically sound and
 mathematically elegant set up whereby the evolution of discontinuous
 and piecewise uniform initial data can be computed, namely, solving a
 RP. Thus, we have carefully set up the initial data of prototype RPs
 in which, either one of the states corresponds to a non-convex
 relativistic thermodynamical state, or the Taub adiabat joining the
 initial data pierces through a region of relativistic non-convexity
 of the EoS. In the former case, we tune the physical state to be
 convex in a classical sense (i.e., the classical fundamental
 derivative evaluated on the initial data is positive). As expected,
 non-convex dynamics can develop structures such as rarefaction
 shocks, compound waves, etc., which are in clear contrast to the wave
 structure that a convex EoS generates.

 The interest of this study is not purely academic and may have far
 reaching implications (beyond the scope of this publication) in
 various fields of Astrophysics (e.g., stellar core collapse of
 massive stars, the merger of compact binaries), Cosmology (e.g., in
 the evolution of the Early Universe), Nuclear Physics (e.g., the
 collision of heavy ions), etc. Precisely, the impact that a
 non-convex EoS may have on the dynamics of stellar core collapse we
 will be the subject of a subsequent paper
 \citep{Ibanez_etal:2017b}. There, we will show that non-convex
 dynamics may leave an imprint on the gravitational wave signature of
 the system, specially if the non-magnetized core rotates fast and
 differentially, which could be observed by future gravitational wave
 detectors or suitable upgrades of the existing ones. We however note
 that the presence of strong (dynamically relevant) magnetic fields in
 a collapsing stellar core may counterbalance the effects of a
 non-convex EoS, both from the mathematical point of view and
 physics-wise. Mathematically, \cite{Sernaetal:2014PoF} in the
 classical MHD case and \cite{Ibanez:2015CQGra..32i5007I} for
 relativistic MHD showed that the fundamental derivative for either
 Newtonian or relativistic, magnetized fluids adds a positively
 defined term to the relativistic fundamental derivative. This term
 grows with increasing magnetic field strength.  Hence, strong
 magnetic fields reduce the ranges of thermodynamic states where the
 fundamental derivative may be negative, i.e., where the system loses
 its convexity. Reinforcing this point, \cite{Sernaetal:2014PoF} show
 numerical examples in which a strong enough magnetic field may revert
 the non-convexity effects induced by two different non-convex
 EoS. From the physical point of view, strong magnetic fields break
 the rotation of massive stellar cores
 \citep[e.g.][]{Yamada_Sawai:2004,Heger_etal:2005,Obergaulinger_Aloy:2017}
 and, therefore, weaken the gravitational wave signature of the system
 \citep{Kotake_etal:2004,Obergaulinger_etal:2006}. These facts
 highlight the need of a deeper numerical study of the potential
 physical effects that non-convex EoS have in the context of
 magneto-rotational core collapse.

\section*{Acknowledgments}
%
We acknowledge support from the European Research Council (grant CAMAP-259276). We also acknowledge support from grants AYA2015-66899-C2-1-P, PROMETEOII/2014-069 and from MINECO, MTM2014-56218-C2-2-P.

%
%
 

\bibliography{nct2}

\label{lastpage}
\end{document}